\documentclass[pra,amsmath,amssymb,notitlepage,twocolumn,nofootinbib,superscriptaddress,longbibliography]{revtex4-2}

\usepackage{amsmath,amsthm}
\usepackage{tabularx,graphicx}
\usepackage{epstopdf}
\usepackage{makecell}
\usepackage{graphicx}
\usepackage{latexsym}
\usepackage{amssymb}
\usepackage{amsmath}
\usepackage{color, colortbl}
\usepackage{mathtools}
\usepackage{psfrag}
\usepackage{bbm}
\usepackage{bm}
\usepackage{titlesec}
\usepackage{dsfont}
\usepackage{feynmp}
\usepackage{slashed}
\usepackage{multirow}
\usepackage{url}
\usepackage{xr}
\usepackage{xcite}
\usepackage{braket}

\newcommand{\onenorm}[1]{\left\| #1 \right\|_1}
\newcommand{\opnorm}[1]{\left\| #1 \right\|}

\newcommand{\tr}{\mathrm {tr}}
\newcommand{\bsp}{\begin{aligned}}
	\newcommand{\esp}{\end{aligned}}

\newcommand{\ie}{{i.e., }}

\newcommand{\mO}{{\hat{O}}}

\newcommand{\supp}{\mathrm{supp}}

\newcommand{\cc}{\mathscr{C}}
\newcommand{\prob}{\mathbb{P}}

\usepackage{xcolor}
\definecolor{darkblue}{rgb}{0.,0.,0.4}
\definecolor{darkred}{rgb}{0.5,0.,0.}
\definecolor{BlueViolet}{RGB}{138,43,226}
\definecolor{SkyBlue}{RGB}{30,144,255}
\definecolor{DarkGreen}{RGB}{0,100,0}

\usepackage[colorlinks=true,linkcolor=darkblue,citecolor=blue,urlcolor=darkred]{hyperref}
\usepackage[normalem]{ulem}

\usepackage{mathrsfs}
\usepackage{todonotes,comment,cleveref}

\newtheorem{corollary}{Corollary}
\newtheorem{theorem}{Theorem}

\newenvironment{theorem_restate}[2]
{\renewcommand{\thetheorem}{\ref{#1}#2}
	\addtocounter{theorem}{-1}
	\begin{theorem}}
	{\end{theorem}}

\usepackage{booktabs}
\usepackage{yfonts}

\usepackage[caption=false]{subfig}

\usepackage{enumitem}

\begin{document}
	
	\title{Lovász Meets Lieb-Schultz-Mattis: Complexity in Approximate Quantum Error Correction}
	
	\newcommand{\PI}{Perimeter Institute for Theoretical Physics, Waterloo, Ontario, Canada N2L 2Y5}
	\newcommand{\UW}{Department of Physics and Astronomy, University of Waterloo, Waterloo, Ontario, Canada N2L 3G1}
	\newcommand{\Dalhousie}{Department of Mathematics and Statistics, Dalhousie University, Halifax, Nova Scotia, Canada B3H 4R2}
	\newcommand{\IBM}{IBM Quantum, IBM Research}
	
	\author{Jinmin Yi}\email[Jinmin Yi:~]{jyi@pitp.ca}
	\affiliation{\PI}
	\affiliation{\UW}
	
	\author{Ruizhi Liu}
	\affiliation{\Dalhousie}
	\affiliation{\PI}
	
	\author{Zhi Li}
	\email[Zhi Li:~]{zli@ibm.com}
	\affiliation{\IBM}

	\begin{abstract}
		
		Approximate quantum error correction (AQEC) provides a versatile framework for both quantum information processing and probing many-body entanglement. We reveal a fundamental tension between the error-correcting power of an AQEC and the hardness of code state preparation. More precisely, through a novel application of the Lovász local lemma, we establish a fundamental trade-off between local indistinguishability and circuit complexity, showing that orthogonal short-range entangled states must be distinguishable via a local operator. These results offer a powerful tool for exploring quantum circuit complexity across diverse settings. As applications, we derive stronger constraints on the complexity of AQEC codes with transversal logical gates and establish strong complexity lower bounds for W state preparation. Our framework also provides a novel perspective for systems with Lieb-Schultz-Mattis type constraints.

	\end{abstract}
	\maketitle

	\section{Introduction} 
	
	Quantum error correction (QEC) is a cornerstone of scalable quantum computation and a powerful lens for exploring fundamental physics \cite{wen1990ground,kitaev2003fault,pastawski2015holographic,almheiri2015bulk}. 
	Central to QEC is the quantum entanglement, which allows information to be encoded nonlocally, rendering it immune to local noise and enabling reliable recovery. 
	The study of entanglement structure in QEC codes has proven exceptionally fruitful, providing a quantitative window into topological phases of matter~\cite{kitaev2006topological,levin2006detecting}, illuminating the emergence of geometry in holographic systems~\cite{Ryu2006,pastawski2015holographic,almheiri2015bulk}, and enabling the resolution of the no low-energy trivial state (NLTS) conjecture~\cite{freedman2013quantum,anshu2023nlts,panteleev2022asymptotically}, to name just a few examples.

	While much of the progress in quantum error correction has focused on exact QEC, where perfect recovery is guaranteed for specific error models, many physically relevant regimes naturally require approximate QEC (AQEC). 
	In these settings, AQEC is not only practically advantageous but often conceptually unavoidable. 
	From a practical perspective, it can achieve higher code rates~\cite{Leung97,liu2025NSA,CGS05:aqec} and enable a broader set of logical operations~\cite{Hayden_2021,faist20,Woods2020continuousgroupsof,PhysRevLett.126.150503,PhysRevResearch.4.023107,Zhou2021newperspectives,liu2022approximate}. 
	Beyond computation, AQEC has also emerged as a versatile toolbox for probing the structure of a wide range of physical systems~\cite{yi_complexity_2024,sang2024approximatequantumerrorcorrecting,BCSB19:aqec,Bentsen_2024}.

	The entanglement structure of (A)QEC codes is often captured by circuit complexity, namely the minimal circuit depth required to prepare a quantum state. 
	It is well established that \emph{all} code states of exact QEC codes must have nontrivial circuit complexity \cite{bravyi2025much}, and similar statements hold for AQEC codes close enough to exact codes \cite{yi_complexity_2024}. 
	However, not all AQECs share this property. 
	As a simple counterexample, consider the AQEC spanned by $\ket{0^n}$ and the W state $\ket{W_n}$ (see \cref{eq:Wstate}), where one of the code states is the trivial product state.
	In fact, similar phenomena appear in many natural physical settings—including translationally invariant spin chains~\cite{BCSB19:aqec}, gapless systems~\cite{yi_complexity_2024}, and holographic models~\cite{Bentsen_2024}—where most states remain highly entangled, yet a small number of states may have low circuit complexity.
	From a technical standpoint, many straightforward extensions of arguments used in exact QEC fail in these regimes, as small errors can accumulate and render the resulting bounds meaningless.

	In this work, we resolve this puzzle through a novel application of the Lovász local lemma (LLL), a cornerstone of probabilistic combinatorics. 
	We show that the presence of multiple low-complexity, mutually orthogonal states is tightly constrained by their local distinguishability. 
	In the context of AQEC, this translates into a fundamental trade-off between error-correcting capability and the circuit complexity required for state preparation. 
	Taken together, these results fill a missing gap in our understanding of the AQEC–complexity interplay and yield a clear ``phase diagram" linking code performance to preparation hardness.

	Beyond the above implications for (A)QEC, our results provide a versatile tool for analyzing circuit complexity across quantum information and condensed-matter physics. 
	We illustrate this framework with three applications.
	%%%%%%%%%%%%%%%%
	First, for covariant AQEC codes, \ie codes admitting transversal logical gates, we prove lower bounds on the code space circuit complexity.
	We show that the presence of transversal gates forces nontrivial complexity throughout the entire code space, even if the AQEC codes are quite far from an exact one.
	%%%%%%%%%%%%%%%%
	Second, for the W state, we derive tight complexity bounds under both geometrically local and all-to-all connectivity constraints.
	%%%%%%%%%%%%%%%
	Finally, we provide an AQEC-based, streamlined proof of Lieb–Schultz–Mattis (LSM) type constraints~\cite{lieb1961two,LSM_oshikawa,LSM_HigherD_Hastings,liu2025entanglementarealawliebschultzmattis,Gioia2021}, covering both the original version and the recent momentum-based variant.

	\section{Preliminaries}\label{sec:preliminaries}

	We begin with a brief overview of approximate quantum error correction (AQEC).
	An $(\!(n,k)\!)$ (approximate or exact) quantum error-correcting code is a $2^k$-dimensional subspace $\mathfrak{C} \subset (\mathbb{C}^2)^{\otimes n}$ that encodes $k$ logical qubits into $n$ physical qubits. 
	To ensure \emph{exact} recovery after errors in a region $S$, the no-cloning principal requires that the region contains no logical information: the code states should be perfectly indistinguishable on small subsystems. 
	Formally, this is captured by the Knill–Laflamme condition \cite{knill1997theory,bennett1996mixed}:
	\begin{equation}\label{eq:KL}
		\tr_{\overline{S}} \big( \ket{\psi_1}\bra{\psi_1}-\ket{\psi_2}\bra{\psi_2} \big) =0,
	\end{equation}
	for any two normalized states $\ket{\psi_1}, \ket{\psi_2}\in \mathfrak{C}$.
	Here, $\tr_{\overline{S}}$ denotes tracing out the complement region of $S$.
	This condition in turn guarantees the existence of a recovery map correcting errors supported on $S$.
	
	In the approximate setting, this strict indistinguishability is relaxed to an approximate version.
	This leads naturally to the \emph{subsystem variance}~\cite{yi_complexity_2024}:
	\begin{equation}
		\varepsilon(d) 
		= \max_{|\psi\rangle\in\mathfrak{C}} 
		\max_{|S|\leq d} 
		\onenorm{\tr_{\overline{S}} \big(|\psi\rangle\langle\psi| - \Gamma\big)}\;,
		\label{eq:subsystem-variance}
	\end{equation}
	where $\Gamma=\frac{1}{2^k}\sum_i |\psi_i\rangle\langle\psi_i|$ is the maximally mixed state on $\mathfrak{C}$.
	Here we have maximized over all subsets $S$ of at most $d$ qubits. 
	
	The function $\varepsilon(d)$ serves as a generalized notion of code distance: for an exact code with distance $d^*$, $\epsilon(d)>0$ if and only if $d\geq d^*$.
	From a more operational perspective, small values of $\varepsilon(d)$ indicate that the code has nice error-correcting properties under certain noise models as measured by the channel distance between the noise–recovery process and the identity~\cite{BO10:AKL,Ng10,yi_complexity_2024}.

	Next, we discuss the other key notion in this work: circuit complexity.
	The circuit complexity $\cc(|\psi\rangle)$ of a pure state $|\psi\rangle$ is defined as the minimal depth of a quantum circuit that prepares $|\psi\rangle$ from the product state $|0^n\rangle$.
	We consider two natural settings for the allowed gate connectivity: the \emph{geometrically local} complexity, where gates act only on neighboring qubits in a spatial lattice, and the \emph{all-to-all} complexity, where gates may act on any pair of qubits.
	(Generalizations to arbitrary connectivity are also straightforward.)
	States with $\cc(|\psi\rangle) = O(1)$ independent of system size $n$ are termed short-range entangled (SRE), while those requiring circuit depth that grows with $n$ are long-range entangled (LRE).
	
	To account for approximate state preparation, we also define a robust notion of complexity:
	\begin{equation} 
		\cc^\delta(|\psi\rangle) \coloneqq \min\{ \cc(|\phi\rangle) : \||\phi\rangle\langle\phi| - |\psi\rangle\langle\psi|\|_1 \le \delta \}, 
	\end{equation}
	where we minimize over all states $|\phi\rangle$ that approximate $|\psi\rangle$ up to an allowed error $\delta$.

	For convenience, we define a \emph{light-cone function} $f(t)$ as the maximum number of qubits that can be influenced by a quantum circuit of depth $t$, given a specified connectivity constraint.
	For example, we have
	\begin{equation}
		f(t) \leq  
		\begin{cases}
			2^t,&~~\text{all-to-all 2-body connectivity}\\
			(2t+1)^D,&~~\text{$D$-dimensional square lattice} 
		\end{cases}.
	\end{equation}

	Finally, let us introduce the Lovász local lemma (LLL).
	The LLL provides conditions under which a collection of events can be avoided simultaneously.
	As a motivation, recall the elementary fact that if $n$ events are independent and each occurs with probability at most $p<1$, then the probability that none of them occurs is at least $(1-p)^n>0.$
	The LLL slightly relaxes the independence assumption at the cost of requiring $p$ to be small. 
	Specifically, suppose there are $n$ events, each of which is independent of all but at most $d$ others. 
	If $p$ satisfies $(d+1)ep < 1$ (here and throughout the paper, $e=\exp(1)$),
	then the probability that none of the $n$ events occurs remains strictly positive:
	\begin{equation}
		\prob[\text{no event occurs}] > (1-ep)^n. 
	\end{equation}

	\section{Distinguishability-Complexity trade-off}\label{sec:main-results}

	Our main theoretical contribution is a fundamental trade-off between approximate local indistinguishability and circuit complexity.
	In the setting of AQEC, this translates into a tension between the error-correcting capability of a code and the circuit complexity to prepare the code.

	\subsection{Distinguishing SRE states}
	
	We begin with our main theorem, which states that nearly orthogonal SRE states can be distinguished by local observables.
	
	\begin{theorem}
		\label{thm:local-distinguishability}
		Let $|\psi_{1,2}\rangle$ be two $n$-qubit states that are almost orthogonal, \ie $|\langle\psi_1|\psi_2\rangle|<\delta$. Suppose $\cc(|\psi_{1,2}\rangle)\leq t$, then there exists a distinguishing operator $\mO$, with $\opnorm{\mO}= 1$ and $|\mathrm{supp}(\mO)| \leq f(t)$, such that
		\begin{equation}
			|\langle\psi_1|\mO|\psi_1\rangle-\langle\psi_2|\mO|\psi_2\rangle|>
			\frac{2}{e}\min
			\big\{1-\delta^{\frac{2}{n}},\frac{1}{f(4t)}\big\}.
			\label{eq:distinguishability-bound}
		\end{equation}
	\end{theorem}

	Here and in the following, $f(t)$ denotes the lightcone function for a given connectivity constraint.
	The operator $\mO$ also inherits the locality structure defining $\cc(\cdot)$ and $f(\cdot)$: 
	in the geometrically local scenario, $\mO$ is supported on a finite lattice region.

	The theorem also extends to mixed states and to cases where one of the states exhibits short-range correlations without being short-range entangled, such as normal matrix product states (MPS) or unique ground states of commuting-projector Hamiltonians. 
	Further details and generalizations are provided in the SM \cite{supp}.

	\begin{proof}
		We now sketch the core intuition (see SM \cite{supp} for the full proof). 
		Given the unitary circuit $U_{2}$ preparing the state $|\psi_{2}\rangle=U_{2}|0^n\rangle$, 
		distinguishing $\ket{\psi_{1}}$ from $\ket{\psi_{2}}$ is equivalent to distinguishing $|\phi\rangle = U_2^\dagger \ket{\psi_1}$ from $|0^n\rangle$.
		Denoting $P_i = |1_i\rangle\langle 1_i|$ on each site $i$, we consider the probability distribution obtained by simultaneously measuring $\{P_i\}$ on $|\phi\rangle$. 
		Define events $E_i$ as “the $i$-th measurement yields~1.” 
		Due to the short-range entanglement nature of $\ket{\phi}$, the dependency among events $\{E_i\}$ only exists for nearby events. 
		
		We claim that at least one $P_i$ can serve as a local observable distinguishing $|\phi\rangle$ from $|0^n\rangle$. 
		Assume the contrary, namely that $|\bra{\phi} P_i \ket{\phi}|$ is small for every~$i$. 
		Consequently, the Lovász local lemma ensures a positive lower bound on the probability that none of $E_i$ occur, which is exactly given by the overlap between $\ket{\psi_1}$ and $\ket{\psi_2}$:
		\begin{equation}
			\prob [\text{none of $E_i$ occurs}] = |\langle \phi | 0^n\rangle|^2=|\braket{\psi_1|\psi_2}|^2.
		\end{equation}
		On the other hand, we have $|\langle \phi | 0^n\rangle|^2 \leq \delta^2$ by setup.
		\cref{eq:distinguishability-bound} is then proved by comparing $\delta^2$ with the LLL bound.
	\end{proof}

	\subsection{Complexity of AQEC code states}

	A natural setting where local (in)distinguishability arises is approximate quantum error correction (AQEC). 
	Here, the error-correction capability is directly tied to the subsystem variance, which quantifies the local distinguishability of code states. 
	Theorem~\ref{thm:local-distinguishability} therefore yields a trade-off between error-correction performance and the circuit complexity required to prepare the code.

	\begin{theorem_restate}{thm:local-distinguishability}{$'$}
		Given an AQEC code with two orthogonal code states $|\psi_1\rangle$ and $|\psi_2\rangle$, 
		if $\cc(|\psi_{1,2}\rangle) \leq t$, then
		\begin{equation}
			\varepsilon(f(t))>\frac{1}{ef(4t)}.
		\end{equation}
	\end{theorem_restate}

	This result follows directly from \cref{thm:local-distinguishability}.
	Note that the subsystem variance upper bounds the distinguishability for any operator $\mO$ such that $\opnorm{\mO}=1$:
	\begin{equation}
		|\langle\psi_1|\mO|\psi_1\rangle-\langle\psi_2|\mO|\psi_2\rangle|\leq2\varepsilon(|\supp(\mO)|).
	\end{equation}
	Together with \cref{eq:distinguishability-bound}, the existence of two orthogonal low-complexity codewords immediately implies a nonzero lower bound on the subsystem variance, concluding \cref{thm:local-distinguishability}$'$.
	We note that while we focused on exact complexity for brevity, our results extend naturally to the approximate complexities as well, see SM \cite{supp}.

	Contrapositively, for any two orthogonal code states in an AQEC with small subsystem variance, at least one must exhibit large circuit complexity.
	From a technical standpoint, this provides a practical method for proving circuit-complexity lower bounds:
	If a state is known to be orthogonal and locally indistinguishable from an auxiliary SRE state, then it necessarily belongs to the LRE class.

	As a special case, the above theorem shows that AQEC with diverging effective code distances forbids constant depth isometric encoders: if the subsystem variance satisfies $\varepsilon(d)=o(1)$ for any constant $d$, then the code space cannot be prepared by a single finite-depth encoder. 
	Related statements for independent noise can also be found in Ref.~\cite{liu2025approximatequantumerrorcorrection}.

	\section{Applications}\label{sec:applications}

	\subsection{Complexity of covariant codes}
	
	In this subsection, we consider codes that admit a group of transversal logical gates, \ie logical gates realizable as tensor products of single-qubit unitaries $U_L=\bigotimes_{i=1}^n U_i$. 
	Such codes are sometimes termed covariant codes.
	Due to the Eastin--Knill theorem~\cite{Eastin_2009}, transversal universal gate sets are incompatible with exact error correction.
	However, we may consider universal gate sets in approximate codes, or discrete logical groups in exact or approximate codes.

	The key insight is that when transversal gates are sufficiently abundant, they can rotate any code state to be nearly orthogonal while keeping the circuit complexity invariant.
	Our main theorems then give a trade-off between the local indistinguishability and the circuit complexity of all code states.

	If a code admits a universal transversal gate set, then any logical unitary $U_L$ can be approximated to arbitrary accuracy using transversal gates, since transversality is preserved under gate composition.
	We then have the following corollary:
	\begin{corollary}
		Given an $(\!(n,k)\!)$ covariant code with a universal transversal gate set, if a number $t$ satisfies
		\begin{equation}\label{eq:covariant_code_condition}
			\varepsilon(f(t))\leq\frac{1}{ef(4t)}\;,
		\end{equation}
		then every code state $|\psi\rangle$ satisfies $\cc(|\psi\rangle)>t$.
	\end{corollary}

	One of the most important discrete groups in quantum computing is the Clifford group.
	Owing to its unitary 2-design property, Clifford gates can map any code state to another that is nearly orthogonal, and hence the same arguments apply, yielding:
	\begin{corollary}\label{thm:Clifford}
		Given an $(\!(n,k)\!)$ code with transversal Clifford logical gates, if 
		\begin{equation}\label{eq:clifford_code_condition}
			\varepsilon(f(t))\leq \frac{1}{e}\min\left\{1-2^{-\frac{k}{n}},\frac{1}{f(4t)}\right\}\;,
		\end{equation}
		every code state $|\psi\rangle$ satisfies $\cc(|\psi\rangle)>t$.
	\end{corollary}

	Both results provide nontrivial lower bounds under notably weak assumptions. 
	For many AQECs, one typically expects $\varepsilon(d)\sim d/n$~\cite{faist20}, a regime not captured by previous approaches~\cite{yi_complexity_2024}.
	Nevertheless, as long as there exists a diverging parameter $x$ (e.g., $x\sim\log n$) such that $\varepsilon(x(n))=O(1/\mathrm{poly}(x(n)))$, \cref{eq:covariant_code_condition} and \cref{eq:clifford_code_condition} will hold, yielding meaningful consequences\footnote{Here, the precise polynomial degree depends on the relation between $f(4t)$ and $f(t)$, determined by the underlying connectivity. 
		For \cref{thm:Clifford}, we also need a moderate coding rate $k/n$, depending on the choice of $x$. See SM \cite{supp} for more details.}.
	Ultimately, transversal gates act as an additional structure that imposes stronger constraints on entanglements.

	\subsection{W state preparation}
	
	Our results provide a powerful method for probing the circuit complexity of quantum states.
	As a concrete illustration, consider the $n$-qubit W state
	\begin{equation}\label{eq:Wstate}
		|W_n\rangle \;=\; \frac{1}{\sqrt{n}}\sum_{i=1}^n |0\cdots 01_i0\cdots 0\rangle\;.
	\end{equation}
	Applying our framework yields the following bounds for both geometrically local and all-to-all circuits:
	\begin{corollary}
		For $\delta<1/10$, the geometrically local circuit complexity of $|W_n\rangle$ on a 1D chain is 
		\begin{equation}
			\cc^\delta(|W_n\rangle)=\Omega(n).
		\end{equation}
		For $\delta<1/n^\alpha$ with $\alpha>1/2$, the all-to-all circuit complexity of $|W_n\rangle$ is 
		\begin{equation}
			\cc^\delta(|W_n\rangle)=\Omega(\log n).
		\end{equation}
	\end{corollary}
	The key observation is that $|W_n\rangle$ and $|0^n\rangle$ are orthogonal yet nearly locally indistinguishable.
	In fact, they form an orthonormal basis of an AQEC.
	The proof then parallels the argument of \cref{thm:local-distinguishability}, with a slight modification (see SM \cite{supp} for the full proof): we partition the $n$ qubits into patches of suitably chosen size and define the local event ``the patch contains at least one 1.''
	
	The scaling in $n$ in our lower bounds match the complexities of known concrete circuits~\cite{W_LNN,Bartschi_2022,liu2025lowdepthquantumsymmetrization,yuan2025depthefficientquantumcircuitsynthesis} and is therefore optimal\footnote{For the scaling of $\delta$ in the all-to-all case, we improved it to $\delta<n^{-\alpha}$ for any $\alpha>0$ via a different method, see SM \cite{supp}}.
	To our best knowledge, both our lower bounds are beyond known results when $\delta\neq 0$. 
	
	\subsection{Lieb-Schultz-Mattis theorem}
	The trade-off between AQEC and circuit complexity 
	also yields a streamlined route to probing long-range entanglements in anomalous quantum many-body systems. 
	Below, we present an entanglement version of the LSM theorem and a momentum variant. 
	
	We start with the canonical case.
	Consider a 1D system of size $L$ with the lattice translation operator $\hat{T}$ and an on-site $U(1)$ transformation $\otimes_{x=0}^{L-1}\exp(i \theta \hat{q}_x)$ where $\theta\in[0,2\pi)$ and each $\hat{q}_x$ has only integer eigenvalues.
	We denote $\hat{Q}=\sum_{x=0}^{L-1}\hat{q}_x$.
	\begin{corollary}[$U(1)\times T$ LSM]\label{thm:LSM}
		Suppose $\ket{\psi}$ is both translationally invariant
		\begin{align}
			\hat{T}\ket{\psi}\propto \ket{\psi},
		\end{align}
		and $U(1)$ symmetric with non-commensurate charge filling
		\begin{equation}\label{eq:non-commensurate}
			\exp(\frac{2\pi i\hat{Q}}{L})\ket{\psi} = e^{i\alpha} \ket{\psi},~~e^{i\alpha}\neq 1.
		\end{equation}
		Then $\ket{\psi}$ cannot be prepared by finite-depth circuits or finite-time Hamiltonian evolution.
	\end{corollary}

	Here, we only show that the geometric local complexity  $\cc(|\psi\rangle)\neq O(1)$, deferring the full proof to the SM \cite{supp}.
	The argument proceeds by constructing a locally indistinguishable partner for $\ket{\psi}$ and then applying \cref{thm:local-distinguishability}.
	Specifically, consider the large gauge transformation
	\begin{equation}
		U:=\exp(\frac{2\pi i}{L}\sum_{x=0}^{L-1}x\,\hat{q}_{x}).
	\end{equation}
	Despite the apparent discontinuity at $L$, the operator is in fact continuous due to the integer eigenvalues of the $\hat{q}_x$.
	It is straightforward to check that $\ket{\psi}$ and $U\ket{\psi}$ are locally indistinguishable up to $O(1/L)$.
	Moreover, it is evident that $\cc(|\psi\rangle)=\cc(U|\psi\rangle)$.
	
	On the other hand, \cref{eq:non-commensurate} together with $U^\dagger \hat{T} U=\exp(-\frac{2\pi i\hat{Q}}{L})\hat{T}$ implies that $\ket{\psi}$ and $U\ket{\psi}$ have distinct lattice momenta and are therefore orthogonal.
	If, contrary to our claim, $\cc(|\psi\rangle)=O(1)$, then \cref{thm:local-distinguishability} would guarantee the existence of a size-$O(1)$ operator that distinguishes them by $\Omega(1)$—contradicting the $O(1/L)$ indistinguishability.

	Recently, interesting progress has been made in understanding the LSM theorem from a momentum-based perspective~\cite{Gioia2021}, which in turn implies the $\cc(|\psi\rangle)\neq O(1)$ part of \cref{thm:LSM}.
	Within our framework, this result can be naturally recovered.
	
	\begin{corollary}[Nonzero momentum implies LRE]
		Let $|\psi\rangle$ be a state on a 1D system of size L. Suppose it is translationally invariant with a non-zero momentum:
		\begin{equation}
			\hat{T}|\psi\rangle=e^{ip}|\psi\rangle,\quad e^{ip}\neq 1,
		\end{equation}
		then the geometric local complexity $\cc(|\psi\rangle)=\Omega(L)$.
	\end{corollary}
	The idea is still to construct a partner state that is short-range correlated, orthogonal, and locally indistinguishable from $|\psi\rangle$.
	It is constructed as follows (see SM \cite{supp} for details).
	Suppose $U$ is a finite-depth unitary circuit that prepares $|\psi\rangle$.
	We cut the circuit, make copies, and tile them together to build a circuit $\tilde U$ on the infinite chain.
	By design, $\tilde U$ would prepare a translation-invariant, short-range entangled state, which admits a translational invariant infinite MPS (iMPS) representation~\cite{fannes1992finitely}. 
	We then truncate this iMPS to a finite ring of length $L$ with periodic boundary conditions, yielding a state $|\psi'\rangle$ on the original system.
	By construction, $|\psi'\rangle$ has zero momentum (hence orthogonal to $\ket{\psi}$) and is locally indistinguishable from $|\psi\rangle$ on $O(L)$ scales. 
	Therefore, by \cref{thm:local-distinguishability}, $|\psi\rangle$ cannot be short-range entangled. 
	
	\section{Discussion and outlook}\label{sec:conclusion}
	
	\begin{figure}[!t]
		\centering
		\includegraphics[width=0.6\linewidth]{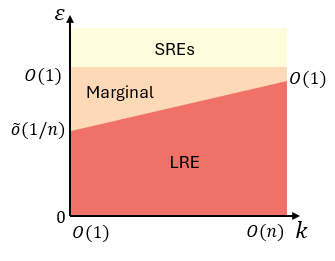}
		\caption{The trade-off between AQEC capability and code state complexity. In the ``LRE" region, the whole code subspace has a complexity lower bound. If the code subspace contains two nearly orthogonal SRE states, the capability will fall into the ``SREs" region. In the marginal region, there can be at most one SRE state in any orthogonal basis.}
		\label{fig:phase_diagram}
	\end{figure}

	In this work, we established a fundamental link between circuit complexity and approximate quantum error correction by analyzing approximate local distinguishability through the lens of the Lovász local lemma. Our results not only resolve a key puzzle in AQEC but also provide broadly applicable insights for quantum information theory and quantum many-body physics.

	This result sharpens the known trade-offs between AQEC capability and circuit complexity, completing a qualitative ``phase diagram" as shown in Fig.~\ref{fig:phase_diagram}. 
	Recent work~\cite{yi_complexity_2024} has shown that when the subsystem variance lies below a $\sim 1/n$ threshold, the entire code space is long-range entangled.
	We show that while a single SRE state may exist in the ``marginal" region, the presence of a second nearly orthogonal SRE state immediately degrades the code performance, forcing the subsystem variance to be $\Omega(1)$. 
	This has immediate consequences for covariant codes, where certain transversal logical gates can generate such nearly orthogonal pairs from a single SRE state. 
	Consequently, such transversal logical gates can enforce code subspace complexity even in the marginal region.

	Beyond the setting of quantum error correction, our framework provides a unifying method for deriving circuit-complexity lower bounds across quantum information and many-body physics. 
	By contrasting a target state with an orthogonal reference state, our distinguishability principle recovers tight preparation bounds for the W state and yields a streamlined, information-theoretic perspective of Lieb–Schultz–Mattis type theorems.
	This approach recasts these classic condensed matter constraints as a consequence of a single principle: the ability of the system to form an AQEC code in the marginal or LRE region.

	Looking ahead, these results highlight AQEC as a unifying language for complexity across diverse domains. 
	Promising directions include extending the framework to many-body systems with discrete anomalous symmetries and to holographic models, probing whether AQEC encodes universal limits in quantum complexity, and developing analogues for mixed-state phases.
	In this way, AQEC transcends its traditional role in quantum information processing and provides a practical framework for understanding complexity and entanglement across quantum science.

	\begin{acknowledgments}
		We thank Sergey Bravyi, Tyler Ellison, David Gosset, Daniel Gottesman, and Chong Wang for valuable discussions. 
		J.Y. thanks Weicheng Ye, Daniel Gottesman, and Zi-Wen Liu; 
		Z.L. thanks Sergey Bravyi, Dongjin Lee, and Beni Yoshida for previous collaborations, which together inspired the present work.
		Research at Perimeter Institute is supported in part by the Government of Canada through the Department of Innovation, Science and Economic Development and by the Province of Ontario through the Ministry of Colleges and Universities.
	\end{acknowledgments}
	
	\bibliography{ref.bib}

\begin{thebibliography}{5}%
\makeatletter
\providecommand \@ifxundefined [1]{%
 \@ifx{#1\undefined}
}%
\providecommand \@ifnum [1]{%
 \ifnum #1\expandafter \@firstoftwo
 \else \expandafter \@secondoftwo
 \fi
}%
\providecommand \@ifx [1]{%
 \ifx #1\expandafter \@firstoftwo
 \else \expandafter \@secondoftwo
 \fi
}%
\providecommand \natexlab [1]{#1}%
\providecommand \enquote  [1]{``#1''}%
\providecommand \bibnamefont  [1]{#1}%
\providecommand \bibfnamefont [1]{#1}%
\providecommand \citenamefont [1]{#1}%
\providecommand \href@noop [0]{\@secondoftwo}%
\providecommand \href [0]{\begingroup \@sanitize@url \@href}%
\providecommand \@href[1]{\@@startlink{#1}\@@href}%
\providecommand \@@href[1]{\endgroup#1\@@endlink}%
\providecommand \@sanitize@url [0]{\catcode `\\12\catcode `\$12\catcode
  `\&12\catcode `\#12\catcode `\^12\catcode `\_12\catcode `\%12\relax}%
\providecommand \@@startlink[1]{}%
\providecommand \@@endlink[0]{}%
\providecommand \url  [0]{\begingroup\@sanitize@url \@url }%
\providecommand \@url [1]{\endgroup\@href {#1}{\urlprefix }}%
\providecommand \urlprefix  [0]{URL }%
\providecommand \Eprint [0]{\href }%
\providecommand \doibase [0]{https://doi.org/}%
\providecommand \selectlanguage [0]{\@gobble}%
\providecommand \bibinfo  [0]{\@secondoftwo}%
\providecommand \bibfield  [0]{\@secondoftwo}%
\providecommand \translation [1]{[#1]}%
\providecommand \BibitemOpen [0]{}%
\providecommand \bibitemStop [0]{}%
\providecommand \bibitemNoStop [0]{.\EOS\space}%
\providecommand \EOS [0]{\spacefactor3000\relax}%
\providecommand \BibitemShut  [1]{\csname bibitem#1\endcsname}%
\let\auto@bib@innerbib\@empty
%</preamble>
\bibitem [{\citenamefont {Cirac}\ \emph {et~al.}(2021)\citenamefont {Cirac},
  \citenamefont {Perez-Garcia}, \citenamefont {Schuch},\ and\ \citenamefont
  {Verstraete}}]{cirac2021matrix}%
  \BibitemOpen
  \bibfield  {author} {\bibinfo {author} {\bibfnamefont {J.~I.}\ \bibnamefont
  {Cirac}}, \bibinfo {author} {\bibfnamefont {D.}~\bibnamefont {Perez-Garcia}},
  \bibinfo {author} {\bibfnamefont {N.}~\bibnamefont {Schuch}},\ and\ \bibinfo
  {author} {\bibfnamefont {F.}~\bibnamefont {Verstraete}},\ }\bibfield  {title}
  {\bibinfo {title} {Matrix product states and projected entangled pair states:
  Concepts, symmetries, theorems},\ }\href@noop {} {\bibfield  {journal}
  {\bibinfo  {journal} {Reviews of Modern Physics}\ }\textbf {\bibinfo {volume}
  {93}},\ \bibinfo {pages} {045003} (\bibinfo {year} {2021})}\BibitemShut
  {NoStop}%
\bibitem [{\citenamefont {Yi}\ \emph {et~al.}(2024)\citenamefont {Yi},
  \citenamefont {Ye}, \citenamefont {Gottesman},\ and\ \citenamefont
  {Liu}}]{yi_complexity_2024}%
  \BibitemOpen
  \bibfield  {author} {\bibinfo {author} {\bibfnamefont {J.}~\bibnamefont
  {Yi}}, \bibinfo {author} {\bibfnamefont {W.}~\bibnamefont {Ye}}, \bibinfo
  {author} {\bibfnamefont {D.}~\bibnamefont {Gottesman}},\ and\ \bibinfo
  {author} {\bibfnamefont {Z.-W.}\ \bibnamefont {Liu}},\ }\bibfield  {title}
  {\bibinfo {title} {Complexity and order in approximate quantum
  error-correcting codes},\ }\bibfield  {journal} {\bibinfo  {journal} {Nature
  Physics}\ }\href {https://doi.org/10.1038/s41567-024-02621-x}
  {10.1038/s41567-024-02621-x} (\bibinfo {year} {2024})\BibitemShut {NoStop}%
\bibitem [{\citenamefont {Dankert}\ \emph {et~al.}(2009)\citenamefont
  {Dankert}, \citenamefont {Cleve}, \citenamefont {Emerson},\ and\
  \citenamefont {Livine}}]{Dankert2009}%
  \BibitemOpen
  \bibfield  {author} {\bibinfo {author} {\bibfnamefont {C.}~\bibnamefont
  {Dankert}}, \bibinfo {author} {\bibfnamefont {R.}~\bibnamefont {Cleve}},
  \bibinfo {author} {\bibfnamefont {J.}~\bibnamefont {Emerson}},\ and\ \bibinfo
  {author} {\bibfnamefont {E.}~\bibnamefont {Livine}},\ }\bibfield  {title}
  {\bibinfo {title} {Exact and approximate unitary 2-designs and their
  application to fidelity estimation},\ }\href
  {https://doi.org/10.1103/PhysRevA.80.012304} {\bibfield  {journal} {\bibinfo
  {journal} {Phys. Rev. A}\ }\textbf {\bibinfo {volume} {80}},\ \bibinfo
  {pages} {012304} (\bibinfo {year} {2009})}\BibitemShut {NoStop}%
\bibitem [{\citenamefont {Haah}\ \emph {et~al.}(2021)\citenamefont {Haah},
  \citenamefont {Hastings}, \citenamefont {Kothari},\ and\ \citenamefont
  {Low}}]{HaahEvolutionCircuit}%
  \BibitemOpen
  \bibfield  {author} {\bibinfo {author} {\bibfnamefont {J.}~\bibnamefont
  {Haah}}, \bibinfo {author} {\bibfnamefont {M.~B.}\ \bibnamefont {Hastings}},
  \bibinfo {author} {\bibfnamefont {R.}~\bibnamefont {Kothari}},\ and\ \bibinfo
  {author} {\bibfnamefont {G.~H.}\ \bibnamefont {Low}},\ }\bibfield  {title}
  {\bibinfo {title} {Quantum algorithm for simulating real time evolution of
  lattice hamiltonians},\ }\href@noop {} {\bibfield  {journal} {\bibinfo
  {journal} {SIAM Journal on Computing}\ }\textbf {\bibinfo {volume} {52}},\
  \bibinfo {pages} {FOCS18} (\bibinfo {year} {2021})}\BibitemShut {NoStop}%
\bibitem [{\citenamefont {Fannes}\ \emph {et~al.}(1992)\citenamefont {Fannes},
  \citenamefont {Nachtergaele},\ and\ \citenamefont
  {Werner}}]{fannes1992finitely}%
  \BibitemOpen
  \bibfield  {author} {\bibinfo {author} {\bibfnamefont {M.}~\bibnamefont
  {Fannes}}, \bibinfo {author} {\bibfnamefont {B.}~\bibnamefont
  {Nachtergaele}},\ and\ \bibinfo {author} {\bibfnamefont {R.~F.}\ \bibnamefont
  {Werner}},\ }\bibfield  {title} {\bibinfo {title} {Finitely correlated states
  on quantum spin chains},\ }\href@noop {} {\bibfield  {journal} {\bibinfo
  {journal} {Communications in mathematical physics}\ }\textbf {\bibinfo
  {volume} {144}},\ \bibinfo {pages} {443} (\bibinfo {year}
  {1992})}\BibitemShut {NoStop}%
\end{thebibliography}%


\begin{thebibliography}{42}%
\makeatletter
\providecommand \@ifxundefined [1]{%
 \@ifx{#1\undefined}
}%
\providecommand \@ifnum [1]{%
 \ifnum #1\expandafter \@firstoftwo
 \else \expandafter \@secondoftwo
 \fi
}%
\providecommand \@ifx [1]{%
 \ifx #1\expandafter \@firstoftwo
 \else \expandafter \@secondoftwo
 \fi
}%
\providecommand \natexlab [1]{#1}%
\providecommand \enquote  [1]{``#1''}%
\providecommand \bibnamefont  [1]{#1}%
\providecommand \bibfnamefont [1]{#1}%
\providecommand \citenamefont [1]{#1}%
\providecommand \href@noop [0]{\@secondoftwo}%
\providecommand \href [0]{\begingroup \@sanitize@url \@href}%
\providecommand \@href[1]{\@@startlink{#1}\@@href}%
\providecommand \@@href[1]{\endgroup#1\@@endlink}%
\providecommand \@sanitize@url [0]{\catcode `\\12\catcode `\$12\catcode
  `\&12\catcode `\#12\catcode `\^12\catcode `\_12\catcode `\%12\relax}%
\providecommand \@@startlink[1]{}%
\providecommand \@@endlink[0]{}%
\providecommand \url  [0]{\begingroup\@sanitize@url \@url }%
\providecommand \@url [1]{\endgroup\@href {#1}{\urlprefix }}%
\providecommand \urlprefix  [0]{URL }%
\providecommand \Eprint [0]{\href }%
\providecommand \doibase [0]{https://doi.org/}%
\providecommand \selectlanguage [0]{\@gobble}%
\providecommand \bibinfo  [0]{\@secondoftwo}%
\providecommand \bibfield  [0]{\@secondoftwo}%
\providecommand \translation [1]{[#1]}%
\providecommand \BibitemOpen [0]{}%
\providecommand \bibitemStop [0]{}%
\providecommand \bibitemNoStop [0]{.\EOS\space}%
\providecommand \EOS [0]{\spacefactor3000\relax}%
\providecommand \BibitemShut  [1]{\csname bibitem#1\endcsname}%
\let\auto@bib@innerbib\@empty
%</preamble>
\bibitem [{\citenamefont {Wen}\ and\ \citenamefont
  {Niu}(1990)}]{wen1990ground}%
  \BibitemOpen
  \bibfield  {author} {\bibinfo {author} {\bibfnamefont {X.-G.}\ \bibnamefont
  {Wen}}\ and\ \bibinfo {author} {\bibfnamefont {Q.}~\bibnamefont {Niu}},\
  }\bibfield  {title} {\bibinfo {title} {Ground-state degeneracy of the
  fractional quantum {H}all states in the presence of a random potential and on
  high-genus {R}iemann surfaces},\ }\href@noop {} {\bibfield  {journal}
  {\bibinfo  {journal} {Physical Review B}\ }\textbf {\bibinfo {volume} {41}},\
  \bibinfo {pages} {9377} (\bibinfo {year} {1990})}\BibitemShut {NoStop}%
\bibitem [{\citenamefont {Kitaev}(2003)}]{kitaev2003fault}%
  \BibitemOpen
  \bibfield  {author} {\bibinfo {author} {\bibfnamefont {A.~Y.}\ \bibnamefont
  {Kitaev}},\ }\bibfield  {title} {\bibinfo {title} {Fault-tolerant quantum
  computation by anyons},\ }\href
  {https://doi.org/10.1016/S0003-4916(02)00018-0} {\bibfield  {journal}
  {\bibinfo  {journal} {Annals of Physics}\ }\textbf {\bibinfo {volume}
  {303}},\ \bibinfo {pages} {2} (\bibinfo {year} {2003})}\BibitemShut {NoStop}%
\bibitem [{\citenamefont {Pastawski}\ \emph {et~al.}(2015)\citenamefont
  {Pastawski}, \citenamefont {Yoshida}, \citenamefont {Harlow},\ and\
  \citenamefont {Preskill}}]{pastawski2015holographic}%
  \BibitemOpen
  \bibfield  {author} {\bibinfo {author} {\bibfnamefont {F.}~\bibnamefont
  {Pastawski}}, \bibinfo {author} {\bibfnamefont {B.}~\bibnamefont {Yoshida}},
  \bibinfo {author} {\bibfnamefont {D.}~\bibnamefont {Harlow}},\ and\ \bibinfo
  {author} {\bibfnamefont {J.}~\bibnamefont {Preskill}},\ }\bibfield  {title}
  {\bibinfo {title} {Holographic quantum error-correcting codes: Toy models for
  the bulk/boundary correspondence},\ }\href@noop {} {\bibfield  {journal}
  {\bibinfo  {journal} {Journal of High Energy Physics}\ }\textbf {\bibinfo
  {volume} {2015}},\ \bibinfo {pages} {1} (\bibinfo {year} {2015})}\BibitemShut
  {NoStop}%
\bibitem [{\citenamefont {Almheiri}\ \emph {et~al.}(2015)\citenamefont
  {Almheiri}, \citenamefont {Dong},\ and\ \citenamefont
  {Harlow}}]{almheiri2015bulk}%
  \BibitemOpen
  \bibfield  {author} {\bibinfo {author} {\bibfnamefont {A.}~\bibnamefont
  {Almheiri}}, \bibinfo {author} {\bibfnamefont {X.}~\bibnamefont {Dong}},\
  and\ \bibinfo {author} {\bibfnamefont {D.}~\bibnamefont {Harlow}},\
  }\bibfield  {title} {\bibinfo {title} {Bulk locality and quantum error
  correction in ads/cft},\ }\href@noop {} {\bibfield  {journal} {\bibinfo
  {journal} {Journal of High Energy Physics}\ }\textbf {\bibinfo {volume}
  {2015}},\ \bibinfo {pages} {1} (\bibinfo {year} {2015})}\BibitemShut
  {NoStop}%
\bibitem [{\citenamefont {Kitaev}\ and\ \citenamefont
  {Preskill}(2006)}]{kitaev2006topological}%
  \BibitemOpen
  \bibfield  {author} {\bibinfo {author} {\bibfnamefont {A.}~\bibnamefont
  {Kitaev}}\ and\ \bibinfo {author} {\bibfnamefont {J.}~\bibnamefont
  {Preskill}},\ }\bibfield  {title} {\bibinfo {title} {Topological entanglement
  entropy},\ }\href@noop {} {\bibfield  {journal} {\bibinfo  {journal}
  {Physical review letters}\ }\textbf {\bibinfo {volume} {96}},\ \bibinfo
  {pages} {110404} (\bibinfo {year} {2006})}\BibitemShut {NoStop}%
\bibitem [{\citenamefont {Levin}\ and\ \citenamefont
  {Wen}(2006)}]{levin2006detecting}%
  \BibitemOpen
  \bibfield  {author} {\bibinfo {author} {\bibfnamefont {M.}~\bibnamefont
  {Levin}}\ and\ \bibinfo {author} {\bibfnamefont {X.-G.}\ \bibnamefont
  {Wen}},\ }\bibfield  {title} {\bibinfo {title} {Detecting topological order
  in a ground state wave function},\ }\href
  {https://doi.org/10.1103/PhysRevLett.96.110405} {\bibfield  {journal}
  {\bibinfo  {journal} {Phys. Rev. Lett.}\ }\textbf {\bibinfo {volume} {96}},\
  \bibinfo {pages} {110405} (\bibinfo {year} {2006})}\BibitemShut {NoStop}%
\bibitem [{\citenamefont {Ryu}\ and\ \citenamefont
  {Takayanagi}(2006)}]{Ryu2006}%
  \BibitemOpen
  \bibfield  {author} {\bibinfo {author} {\bibfnamefont {S.}~\bibnamefont
  {Ryu}}\ and\ \bibinfo {author} {\bibfnamefont {T.}~\bibnamefont
  {Takayanagi}},\ }\bibfield  {title} {\bibinfo {title} {Holographic derivation
  of entanglement entropy from the anti--de sitter space/conformal field theory
  correspondence},\ }\href {https://doi.org/10.1103/PhysRevLett.96.181602}
  {\bibfield  {journal} {\bibinfo  {journal} {Phys. Rev. Lett.}\ }\textbf
  {\bibinfo {volume} {96}},\ \bibinfo {pages} {181602} (\bibinfo {year}
  {2006})}\BibitemShut {NoStop}%
\bibitem [{\citenamefont {Freedman}\ and\ \citenamefont
  {Hastings}(2014)}]{freedman2013quantum}%
  \BibitemOpen
  \bibfield  {author} {\bibinfo {author} {\bibfnamefont {M.~H.}\ \bibnamefont
  {Freedman}}\ and\ \bibinfo {author} {\bibfnamefont {M.~B.}\ \bibnamefont
  {Hastings}},\ }\bibfield  {title} {\bibinfo {title} {Quantum systems on
  non-k-hyperfinite complexes: a generalization of classical statistical
  mechanics on expander graphs},\ }\href@noop {} {\bibfield  {journal}
  {\bibinfo  {journal} {Quantum Info. Comput.}\ }\textbf {\bibinfo {volume}
  {14}},\ \bibinfo {pages} {144–180} (\bibinfo {year} {2014})}\BibitemShut
  {NoStop}%
\bibitem [{\citenamefont {Anshu}\ \emph {et~al.}(2023)\citenamefont {Anshu},
  \citenamefont {Breuckmann},\ and\ \citenamefont {Nirkhe}}]{anshu2023nlts}%
  \BibitemOpen
  \bibfield  {author} {\bibinfo {author} {\bibfnamefont {A.}~\bibnamefont
  {Anshu}}, \bibinfo {author} {\bibfnamefont {N.~P.}\ \bibnamefont
  {Breuckmann}},\ and\ \bibinfo {author} {\bibfnamefont {C.}~\bibnamefont
  {Nirkhe}},\ }\bibfield  {title} {\bibinfo {title} {{NLTS} hamiltonians from
  good quantum codes},\ }in\ \href@noop {} {\emph {\bibinfo {booktitle}
  {Proceedings of the 55th Annual ACM Symposium on Theory of Computing}}}\
  (\bibinfo {year} {2023})\ pp.\ \bibinfo {pages} {1090--1096}\BibitemShut
  {NoStop}%
\bibitem [{\citenamefont {Panteleev}\ and\ \citenamefont
  {Kalachev}(2022)}]{panteleev2022asymptotically}%
  \BibitemOpen
  \bibfield  {author} {\bibinfo {author} {\bibfnamefont {P.}~\bibnamefont
  {Panteleev}}\ and\ \bibinfo {author} {\bibfnamefont {G.}~\bibnamefont
  {Kalachev}},\ }\bibfield  {title} {\bibinfo {title} {Asymptotically good
  quantum and locally testable classical {LDPC} codes},\ }in\ \href@noop {}
  {\emph {\bibinfo {booktitle} {Proceedings of the 54th Annual ACM SIGACT
  Symposium on Theory of Computing}}}\ (\bibinfo {year} {2022})\ pp.\ \bibinfo
  {pages} {375--388}\BibitemShut {NoStop}%
\bibitem [{\citenamefont {Leung}\ \emph {et~al.}(1997)\citenamefont {Leung},
  \citenamefont {Nielsen}, \citenamefont {Chuang},\ and\ \citenamefont
  {Yamamoto}}]{Leung97}%
  \BibitemOpen
  \bibfield  {author} {\bibinfo {author} {\bibfnamefont {D.~W.}\ \bibnamefont
  {Leung}}, \bibinfo {author} {\bibfnamefont {M.~A.}\ \bibnamefont {Nielsen}},
  \bibinfo {author} {\bibfnamefont {I.~L.}\ \bibnamefont {Chuang}},\ and\
  \bibinfo {author} {\bibfnamefont {Y.}~\bibnamefont {Yamamoto}},\ }\bibfield
  {title} {\bibinfo {title} {Approximate quantum error correction can lead to
  better codes},\ }\href {https://doi.org/10.1103/PhysRevA.56.2567} {\bibfield
  {journal} {\bibinfo  {journal} {Phys. Rev. A}\ }\textbf {\bibinfo {volume}
  {56}},\ \bibinfo {pages} {2567} (\bibinfo {year} {1997})}\BibitemShut
  {NoStop}%
\bibitem [{\citenamefont {Liu}\ \emph {et~al.}(2025{\natexlab{a}})\citenamefont
  {Liu}, \citenamefont {Zhou}, \citenamefont {Liu},\ and\ \citenamefont
  {Yi}}]{liu2025NSA}%
  \BibitemOpen
  \bibfield  {author} {\bibinfo {author} {\bibfnamefont {S.}~\bibnamefont
  {Liu}}, \bibinfo {author} {\bibfnamefont {S.}~\bibnamefont {Zhou}}, \bibinfo
  {author} {\bibfnamefont {Z.-W.}\ \bibnamefont {Liu}},\ and\ \bibinfo {author}
  {\bibfnamefont {J.}~\bibnamefont {Yi}},\ }\href
  {https://arxiv.org/abs/2503.11783} {\bibinfo {title} {Noise-strength-adapted
  approximate quantum codes inspired by machine learning}} (\bibinfo {year}
  {2025}{\natexlab{a}}),\ \Eprint {https://arxiv.org/abs/2503.11783}
  {arXiv:2503.11783 [quant-ph]} \BibitemShut {NoStop}%
\bibitem [{\citenamefont {Cr{\'e}peau}\ \emph {et~al.}(2005)\citenamefont
  {Cr{\'e}peau}, \citenamefont {Gottesman},\ and\ \citenamefont
  {Smith}}]{CGS05:aqec}%
  \BibitemOpen
  \bibfield  {author} {\bibinfo {author} {\bibfnamefont {C.}~\bibnamefont
  {Cr{\'e}peau}}, \bibinfo {author} {\bibfnamefont {D.}~\bibnamefont
  {Gottesman}},\ and\ \bibinfo {author} {\bibfnamefont {A.~D.}\ \bibnamefont
  {Smith}},\ }\bibfield  {title} {\bibinfo {title} {Approximate quantum
  error-correcting codes and secret sharing schemes.},\ }in\ \href@noop {}
  {\emph {\bibinfo {booktitle} {Eurocrypt}}},\ Vol.\ \bibinfo {volume} {3494}\
  (\bibinfo {organization} {Springer},\ \bibinfo {year} {2005})\ pp.\ \bibinfo
  {pages} {285--301}\BibitemShut {NoStop}%
\bibitem [{\citenamefont {Hayden}\ \emph {et~al.}(2021)\citenamefont {Hayden},
  \citenamefont {Nezami}, \citenamefont {Popescu},\ and\ \citenamefont
  {Salton}}]{Hayden_2021}%
  \BibitemOpen
  \bibfield  {author} {\bibinfo {author} {\bibfnamefont {P.}~\bibnamefont
  {Hayden}}, \bibinfo {author} {\bibfnamefont {S.}~\bibnamefont {Nezami}},
  \bibinfo {author} {\bibfnamefont {S.}~\bibnamefont {Popescu}},\ and\ \bibinfo
  {author} {\bibfnamefont {G.}~\bibnamefont {Salton}},\ }\bibfield  {title}
  {\bibinfo {title} {Error correction of quantum reference frame information},\
  }\bibfield  {journal} {\bibinfo  {journal} {{PRX} Quantum}\ }\textbf
  {\bibinfo {volume} {2}},\ \href {https://doi.org/10.1103/prxquantum.2.010326}
  {10.1103/prxquantum.2.010326} (\bibinfo {year} {2021})\BibitemShut {NoStop}%
\bibitem [{\citenamefont {Faist}\ \emph {et~al.}(2020)\citenamefont {Faist},
  \citenamefont {Nezami}, \citenamefont {Albert}, \citenamefont {Salton},
  \citenamefont {Pastawski}, \citenamefont {Hayden},\ and\ \citenamefont
  {Preskill}}]{faist20}%
  \BibitemOpen
  \bibfield  {author} {\bibinfo {author} {\bibfnamefont {P.}~\bibnamefont
  {Faist}}, \bibinfo {author} {\bibfnamefont {S.}~\bibnamefont {Nezami}},
  \bibinfo {author} {\bibfnamefont {V.~V.}\ \bibnamefont {Albert}}, \bibinfo
  {author} {\bibfnamefont {G.}~\bibnamefont {Salton}}, \bibinfo {author}
  {\bibfnamefont {F.}~\bibnamefont {Pastawski}}, \bibinfo {author}
  {\bibfnamefont {P.}~\bibnamefont {Hayden}},\ and\ \bibinfo {author}
  {\bibfnamefont {J.}~\bibnamefont {Preskill}},\ }\bibfield  {title} {\bibinfo
  {title} {Continuous symmetries and approximate quantum error correction},\
  }\href {https://doi.org/10.1103/PhysRevX.10.041018} {\bibfield  {journal}
  {\bibinfo  {journal} {Phys. Rev. X}\ }\textbf {\bibinfo {volume} {10}},\
  \bibinfo {pages} {041018} (\bibinfo {year} {2020})}\BibitemShut {NoStop}%
\bibitem [{\citenamefont {Woods}\ and\ \citenamefont
  {Alhambra}(2020)}]{Woods2020continuousgroupsof}%
  \BibitemOpen
  \bibfield  {author} {\bibinfo {author} {\bibfnamefont {M.~P.}\ \bibnamefont
  {Woods}}\ and\ \bibinfo {author} {\bibfnamefont {{\'{A}}.~M.}\ \bibnamefont
  {Alhambra}},\ }\bibfield  {title} {\bibinfo {title} {Continuous groups of
  transversal gates for quantum error correcting codes from finite clock
  reference frames},\ }\href {https://doi.org/10.22331/q-2020-03-23-245}
  {\bibfield  {journal} {\bibinfo  {journal} {{Quantum}}\ }\textbf {\bibinfo
  {volume} {4}},\ \bibinfo {pages} {245} (\bibinfo {year} {2020})}\BibitemShut
  {NoStop}%
\bibitem [{\citenamefont {Kubica}\ and\ \citenamefont
  {Demkowicz-Dobrza\ifmmode~\acute{n}\else
  \'{n}\fi{}ski}(2021)}]{PhysRevLett.126.150503}%
  \BibitemOpen
  \bibfield  {author} {\bibinfo {author} {\bibfnamefont {A.}~\bibnamefont
  {Kubica}}\ and\ \bibinfo {author} {\bibfnamefont {R.}~\bibnamefont
  {Demkowicz-Dobrza\ifmmode~\acute{n}\else \'{n}\fi{}ski}},\ }\bibfield
  {title} {\bibinfo {title} {Using quantum metrological bounds in quantum error
  correction: A simple proof of the approximate eastin-knill theorem},\ }\href
  {https://doi.org/10.1103/PhysRevLett.126.150503} {\bibfield  {journal}
  {\bibinfo  {journal} {Phys. Rev. Lett.}\ }\textbf {\bibinfo {volume} {126}},\
  \bibinfo {pages} {150503} (\bibinfo {year} {2021})}\BibitemShut {NoStop}%
\bibitem [{\citenamefont {Yang}\ \emph {et~al.}(2022)\citenamefont {Yang},
  \citenamefont {Mo}, \citenamefont {Renes}, \citenamefont {Chiribella},\ and\
  \citenamefont {Woods}}]{PhysRevResearch.4.023107}%
  \BibitemOpen
  \bibfield  {author} {\bibinfo {author} {\bibfnamefont {Y.}~\bibnamefont
  {Yang}}, \bibinfo {author} {\bibfnamefont {Y.}~\bibnamefont {Mo}}, \bibinfo
  {author} {\bibfnamefont {J.~M.}\ \bibnamefont {Renes}}, \bibinfo {author}
  {\bibfnamefont {G.}~\bibnamefont {Chiribella}},\ and\ \bibinfo {author}
  {\bibfnamefont {M.~P.}\ \bibnamefont {Woods}},\ }\bibfield  {title} {\bibinfo
  {title} {Optimal universal quantum error correction via bounded reference
  frames},\ }\href {https://doi.org/10.1103/PhysRevResearch.4.023107}
  {\bibfield  {journal} {\bibinfo  {journal} {Phys. Rev. Res.}\ }\textbf
  {\bibinfo {volume} {4}},\ \bibinfo {pages} {023107} (\bibinfo {year}
  {2022})}\BibitemShut {NoStop}%
\bibitem [{\citenamefont {Zhou}\ \emph {et~al.}(2021)\citenamefont {Zhou},
  \citenamefont {Liu},\ and\ \citenamefont {Jiang}}]{Zhou2021newperspectives}%
  \BibitemOpen
  \bibfield  {author} {\bibinfo {author} {\bibfnamefont {S.}~\bibnamefont
  {Zhou}}, \bibinfo {author} {\bibfnamefont {Z.-W.}\ \bibnamefont {Liu}},\ and\
  \bibinfo {author} {\bibfnamefont {L.}~\bibnamefont {Jiang}},\ }\bibfield
  {title} {\bibinfo {title} {New perspectives on covariant quantum error
  correction},\ }\href {https://doi.org/10.22331/q-2021-08-09-521} {\bibfield
  {journal} {\bibinfo  {journal} {{Quantum}}\ }\textbf {\bibinfo {volume}
  {5}},\ \bibinfo {pages} {521} (\bibinfo {year} {2021})}\BibitemShut {NoStop}%
\bibitem [{\citenamefont {Liu}\ and\ \citenamefont
  {Zhou}(2023)}]{liu2022approximate}%
  \BibitemOpen
  \bibfield  {author} {\bibinfo {author} {\bibfnamefont {Z.-W.}\ \bibnamefont
  {Liu}}\ and\ \bibinfo {author} {\bibfnamefont {S.}~\bibnamefont {Zhou}},\
  }\bibfield  {title} {\bibinfo {title} {Approximate symmetries and quantum
  error correction},\ }\href {http://dx.doi.org/10.1038/s41534-023-00788-4}
  {\bibfield  {journal} {\bibinfo  {journal} {npj Quantum Information}\
  }\textbf {\bibinfo {volume} {9}} (\bibinfo {year} {2023})}\BibitemShut
  {NoStop}%
\bibitem [{\citenamefont {Yi}\ \emph {et~al.}(2024)\citenamefont {Yi},
  \citenamefont {Ye}, \citenamefont {Gottesman},\ and\ \citenamefont
  {Liu}}]{yi_complexity_2024}%
  \BibitemOpen
  \bibfield  {author} {\bibinfo {author} {\bibfnamefont {J.}~\bibnamefont
  {Yi}}, \bibinfo {author} {\bibfnamefont {W.}~\bibnamefont {Ye}}, \bibinfo
  {author} {\bibfnamefont {D.}~\bibnamefont {Gottesman}},\ and\ \bibinfo
  {author} {\bibfnamefont {Z.-W.}\ \bibnamefont {Liu}},\ }\bibfield  {title}
  {\bibinfo {title} {Complexity and order in approximate quantum
  error-correcting codes},\ }\bibfield  {journal} {\bibinfo  {journal} {Nature
  Physics}\ }\href {https://doi.org/10.1038/s41567-024-02621-x}
  {10.1038/s41567-024-02621-x} (\bibinfo {year} {2024})\BibitemShut {NoStop}%
\bibitem [{\citenamefont {Sang}\ \emph {et~al.}(2024)\citenamefont {Sang},
  \citenamefont {Hsieh},\ and\ \citenamefont
  {Zou}}]{sang2024approximatequantumerrorcorrecting}%
  \BibitemOpen
  \bibfield  {author} {\bibinfo {author} {\bibfnamefont {S.}~\bibnamefont
  {Sang}}, \bibinfo {author} {\bibfnamefont {T.~H.}\ \bibnamefont {Hsieh}},\
  and\ \bibinfo {author} {\bibfnamefont {Y.}~\bibnamefont {Zou}},\ }\href
  {https://arxiv.org/abs/2406.09555} {\bibinfo {title} {Approximate quantum
  error correcting codes from conformal field theory}} (\bibinfo {year}
  {2024}),\ \Eprint {https://arxiv.org/abs/2406.09555} {arXiv:2406.09555
  [quant-ph]} \BibitemShut {NoStop}%
\bibitem [{\citenamefont {Brand\~ao}\ \emph {et~al.}(2019)\citenamefont
  {Brand\~ao}, \citenamefont {Crosson}, \citenamefont {\ifmmode
  \mbox{\c{S}}\else \c{S}\fi{}ahino\ifmmode~\breve{g}\else \u{g}\fi{}lu},\ and\
  \citenamefont {Bowen}}]{BCSB19:aqec}%
  \BibitemOpen
  \bibfield  {author} {\bibinfo {author} {\bibfnamefont {F.~G. S.~L.}\
  \bibnamefont {Brand\~ao}}, \bibinfo {author} {\bibfnamefont {E.}~\bibnamefont
  {Crosson}}, \bibinfo {author} {\bibfnamefont {M.~B.}\ \bibnamefont {\ifmmode
  \mbox{\c{S}}\else \c{S}\fi{}ahino\ifmmode~\breve{g}\else \u{g}\fi{}lu}},\
  and\ \bibinfo {author} {\bibfnamefont {J.}~\bibnamefont {Bowen}},\ }\bibfield
   {title} {\bibinfo {title} {Quantum error correcting codes in eigenstates of
  translation-invariant spin chains},\ }\href
  {https://doi.org/10.1103/PhysRevLett.123.110502} {\bibfield  {journal}
  {\bibinfo  {journal} {Phys. Rev. Lett.}\ }\textbf {\bibinfo {volume} {123}},\
  \bibinfo {pages} {110502} (\bibinfo {year} {2019})}\BibitemShut {NoStop}%
\bibitem [{\citenamefont {Bentsen}\ \emph {et~al.}(2024)\citenamefont
  {Bentsen}, \citenamefont {Nguyen},\ and\ \citenamefont
  {Swingle}}]{Bentsen_2024}%
  \BibitemOpen
  \bibfield  {author} {\bibinfo {author} {\bibfnamefont {G.}~\bibnamefont
  {Bentsen}}, \bibinfo {author} {\bibfnamefont {P.}~\bibnamefont {Nguyen}},\
  and\ \bibinfo {author} {\bibfnamefont {B.}~\bibnamefont {Swingle}},\
  }\bibfield  {title} {\bibinfo {title} {Approximate quantum codes from long
  wormholes},\ }\href {https://doi.org/10.22331/q-2024-08-14-1439} {\bibfield
  {journal} {\bibinfo  {journal} {Quantum}\ }\textbf {\bibinfo {volume} {8}},\
  \bibinfo {pages} {1439} (\bibinfo {year} {2024})}\BibitemShut {NoStop}%
\bibitem [{\citenamefont {Bravyi}\ \emph {et~al.}(2025)\citenamefont {Bravyi},
  \citenamefont {Lee}, \citenamefont {Li},\ and\ \citenamefont
  {Yoshida}}]{bravyi2025much}%
  \BibitemOpen
  \bibfield  {author} {\bibinfo {author} {\bibfnamefont {S.}~\bibnamefont
  {Bravyi}}, \bibinfo {author} {\bibfnamefont {D.}~\bibnamefont {Lee}},
  \bibinfo {author} {\bibfnamefont {Z.}~\bibnamefont {Li}},\ and\ \bibinfo
  {author} {\bibfnamefont {B.}~\bibnamefont {Yoshida}},\ }\bibfield  {title}
  {\bibinfo {title} {How much entanglement is needed for quantum error
  correction?},\ }\href@noop {} {\bibfield  {journal} {\bibinfo  {journal}
  {Physical Review Letters}\ }\textbf {\bibinfo {volume} {134}},\ \bibinfo
  {pages} {210602} (\bibinfo {year} {2025})}\BibitemShut {NoStop}%
\bibitem [{\citenamefont {Lieb}\ \emph {et~al.}(1961)\citenamefont {Lieb},
  \citenamefont {Schultz},\ and\ \citenamefont {Mattis}}]{lieb1961two}%
  \BibitemOpen
  \bibfield  {author} {\bibinfo {author} {\bibfnamefont {E.}~\bibnamefont
  {Lieb}}, \bibinfo {author} {\bibfnamefont {T.}~\bibnamefont {Schultz}},\ and\
  \bibinfo {author} {\bibfnamefont {D.}~\bibnamefont {Mattis}},\ }\bibfield
  {title} {\bibinfo {title} {Two soluble models of an antiferromagnetic
  chain},\ }\href
  {https://www.sciencedirect.com/science/article/abs/pii/0003491661901154}
  {\bibfield  {journal} {\bibinfo  {journal} {Annals of Physics}\ }\textbf
  {\bibinfo {volume} {16}},\ \bibinfo {pages} {407} (\bibinfo {year}
  {1961})}\BibitemShut {NoStop}%
\bibitem [{\citenamefont {{Oshikawa}}(2000)}]{LSM_oshikawa}%
  \BibitemOpen
  \bibfield  {author} {\bibinfo {author} {\bibfnamefont {M.}~\bibnamefont
  {{Oshikawa}}},\ }\bibfield  {title} {\bibinfo {title} {{Commensurability,
  Excitation Gap, and Topology in Quantum Many-Particle Systems on a Periodic
  Lattice}},\ }\href {https://doi.org/10.1103/PhysRevLett.84.1535} {\bibfield
  {journal} {\bibinfo  {journal} {\prl}\ }\textbf {\bibinfo {volume} {84}},\
  \bibinfo {pages} {1535} (\bibinfo {year} {2000})},\ \Eprint
  {https://arxiv.org/abs/cond-mat/9911137} {arXiv:cond-mat/9911137
  [cond-mat.str-el]} \BibitemShut {NoStop}%
\bibitem [{\citenamefont {Hastings}(2004)}]{LSM_HigherD_Hastings}%
  \BibitemOpen
  \bibfield  {author} {\bibinfo {author} {\bibfnamefont {M.~B.}\ \bibnamefont
  {Hastings}},\ }\bibfield  {title} {\bibinfo {title} {Lieb-schultz-mattis in
  higher dimensions},\ }\href {https://doi.org/10.1103/PhysRevB.69.104431}
  {\bibfield  {journal} {\bibinfo  {journal} {Phys. Rev. B}\ }\textbf {\bibinfo
  {volume} {69}},\ \bibinfo {pages} {104431} (\bibinfo {year}
  {2004})}\BibitemShut {NoStop}%
\bibitem [{\citenamefont {Liu}\ \emph {et~al.}(2025{\natexlab{b}})\citenamefont
  {Liu}, \citenamefont {Yi}, \citenamefont {Zhou},\ and\ \citenamefont
  {Zou}}]{liu2025entanglementarealawliebschultzmattis}%
  \BibitemOpen
  \bibfield  {author} {\bibinfo {author} {\bibfnamefont {R.}~\bibnamefont
  {Liu}}, \bibinfo {author} {\bibfnamefont {J.}~\bibnamefont {Yi}}, \bibinfo
  {author} {\bibfnamefont {S.}~\bibnamefont {Zhou}},\ and\ \bibinfo {author}
  {\bibfnamefont {L.}~\bibnamefont {Zou}},\ }\href
  {https://arxiv.org/abs/2405.14929} {\bibinfo {title} {Entanglement area law
  and lieb-schultz-mattis theorem in long-range interacting systems, and
  symmetry-enforced long-range entanglement}} (\bibinfo {year}
  {2025}{\natexlab{b}}),\ \Eprint {https://arxiv.org/abs/2405.14929}
  {arXiv:2405.14929 [cond-mat.str-el]} \BibitemShut {NoStop}%
\bibitem [{\citenamefont {Gioia}\ and\ \citenamefont {Wang}(2022)}]{Gioia2021}%
  \BibitemOpen
  \bibfield  {author} {\bibinfo {author} {\bibfnamefont {L.}~\bibnamefont
  {Gioia}}\ and\ \bibinfo {author} {\bibfnamefont {C.}~\bibnamefont {Wang}},\
  }\bibfield  {title} {\bibinfo {title} {Nonzero momentum requires long-range
  entanglement},\ }\href {https://doi.org/10.1103/PhysRevX.12.031007}
  {\bibfield  {journal} {\bibinfo  {journal} {Phys. Rev. X}\ }\textbf {\bibinfo
  {volume} {12}},\ \bibinfo {pages} {031007} (\bibinfo {year}
  {2022})}\BibitemShut {NoStop}%
\bibitem [{\citenamefont {Knill}\ and\ \citenamefont
  {Laflamme}(1997)}]{knill1997theory}%
  \BibitemOpen
  \bibfield  {author} {\bibinfo {author} {\bibfnamefont {E.}~\bibnamefont
  {Knill}}\ and\ \bibinfo {author} {\bibfnamefont {R.}~\bibnamefont
  {Laflamme}},\ }\bibfield  {title} {\bibinfo {title} {Theory of quantum
  error-correcting codes},\ }\href@noop {} {\bibfield  {journal} {\bibinfo
  {journal} {Physical Review A}\ }\textbf {\bibinfo {volume} {55}},\ \bibinfo
  {pages} {900} (\bibinfo {year} {1997})}\BibitemShut {NoStop}%
\bibitem [{\citenamefont {Bennett}\ \emph {et~al.}(1996)\citenamefont
  {Bennett}, \citenamefont {DiVincenzo}, \citenamefont {Smolin},\ and\
  \citenamefont {Wootters}}]{bennett1996mixed}%
  \BibitemOpen
  \bibfield  {author} {\bibinfo {author} {\bibfnamefont {C.~H.}\ \bibnamefont
  {Bennett}}, \bibinfo {author} {\bibfnamefont {D.~P.}\ \bibnamefont
  {DiVincenzo}}, \bibinfo {author} {\bibfnamefont {J.~A.}\ \bibnamefont
  {Smolin}},\ and\ \bibinfo {author} {\bibfnamefont {W.~K.}\ \bibnamefont
  {Wootters}},\ }\bibfield  {title} {\bibinfo {title} {Mixed-state entanglement
  and quantum error correction},\ }\href@noop {} {\bibfield  {journal}
  {\bibinfo  {journal} {Physical Review A}\ }\textbf {\bibinfo {volume} {54}},\
  \bibinfo {pages} {3824} (\bibinfo {year} {1996})}\BibitemShut {NoStop}%
\bibitem [{\citenamefont {{B{\'e}ny}}\ and\ \citenamefont
  {{Oreshkov}}(2010)}]{BO10:AKL}%
  \BibitemOpen
  \bibfield  {author} {\bibinfo {author} {\bibfnamefont {C.}~\bibnamefont
  {{B{\'e}ny}}}\ and\ \bibinfo {author} {\bibfnamefont {O.}~\bibnamefont
  {{Oreshkov}}},\ }\bibfield  {title} {\bibinfo {title} {{General Conditions
  for Approximate Quantum Error Correction and Near-Optimal Recovery
  Channels}},\ }\href {https://doi.org/10.1103/PhysRevLett.104.120501}
  {\bibfield  {journal} {\bibinfo  {journal} {\prl}\ }\textbf {\bibinfo
  {volume} {104}},\ \bibinfo {eid} {120501} (\bibinfo {year}
  {2010})}\BibitemShut {NoStop}%
\bibitem [{\citenamefont {Ng}\ and\ \citenamefont {Mandayam}(2010)}]{Ng10}%
  \BibitemOpen
  \bibfield  {author} {\bibinfo {author} {\bibfnamefont {H.~K.}\ \bibnamefont
  {Ng}}\ and\ \bibinfo {author} {\bibfnamefont {P.}~\bibnamefont {Mandayam}},\
  }\bibfield  {title} {\bibinfo {title} {Simple approach to approximate quantum
  error correction based on the transpose channel},\ }\href
  {https://doi.org/10.1103/PhysRevA.81.062342} {\bibfield  {journal} {\bibinfo
  {journal} {Phys. Rev. A}\ }\textbf {\bibinfo {volume} {81}},\ \bibinfo
  {pages} {062342} (\bibinfo {year} {2010})}\BibitemShut {NoStop}%
\bibitem [{sup()}]{supp}%
  \BibitemOpen
  \href@noop {} {}\bibinfo {note} {Supplemental Material}\BibitemShut {NoStop}%
\bibitem [{\citenamefont {Liu}\ \emph {et~al.}(2025{\natexlab{c}})\citenamefont
  {Liu}, \citenamefont {Du}, \citenamefont {Liu},\ and\ \citenamefont
  {Ma}}]{liu2025approximatequantumerrorcorrection}%
  \BibitemOpen
  \bibfield  {author} {\bibinfo {author} {\bibfnamefont {G.}~\bibnamefont
  {Liu}}, \bibinfo {author} {\bibfnamefont {Z.}~\bibnamefont {Du}}, \bibinfo
  {author} {\bibfnamefont {Z.-W.}\ \bibnamefont {Liu}},\ and\ \bibinfo {author}
  {\bibfnamefont {X.}~\bibnamefont {Ma}},\ }\href
  {https://arxiv.org/abs/2503.17759} {\bibinfo {title} {Approximate quantum
  error correction with 1d log-depth circuits}} (\bibinfo {year}
  {2025}{\natexlab{c}}),\ \Eprint {https://arxiv.org/abs/2503.17759}
  {arXiv:2503.17759 [quant-ph]} \BibitemShut {NoStop}%
\bibitem [{\citenamefont {Eastin}\ and\ \citenamefont
  {Knill}(2009)}]{Eastin_2009}%
  \BibitemOpen
  \bibfield  {author} {\bibinfo {author} {\bibfnamefont {B.}~\bibnamefont
  {Eastin}}\ and\ \bibinfo {author} {\bibfnamefont {E.}~\bibnamefont {Knill}},\
  }\bibfield  {title} {\bibinfo {title} {Restrictions on transversal encoded
  quantum gate sets},\ }\bibfield  {journal} {\bibinfo  {journal} {Physical
  Review Letters}\ }\textbf {\bibinfo {volume} {102}},\ \href
  {https://doi.org/10.1103/physrevlett.102.110502}
  {10.1103/physrevlett.102.110502} (\bibinfo {year} {2009})\BibitemShut
  {NoStop}%
\bibitem [{\citenamefont {B{\"a}rtschi}\ and\ \citenamefont
  {Eidenbenz}(2019)}]{W_LNN}%
  \BibitemOpen
  \bibfield  {author} {\bibinfo {author} {\bibfnamefont {A.}~\bibnamefont
  {B{\"a}rtschi}}\ and\ \bibinfo {author} {\bibfnamefont {S.}~\bibnamefont
  {Eidenbenz}},\ }\bibfield  {title} {\bibinfo {title} {Deterministic
  preparation of dicke states},\ }in\ \href@noop {} {\emph {\bibinfo
  {booktitle} {International Symposium on Fundamentals of Computation
  Theory}}}\ (\bibinfo {organization} {Springer},\ \bibinfo {year} {2019})\
  pp.\ \bibinfo {pages} {126--139}\BibitemShut {NoStop}%
\bibitem [{\citenamefont {Bartschi}\ and\ \citenamefont
  {Eidenbenz}(2022)}]{Bartschi_2022}%
  \BibitemOpen
  \bibfield  {author} {\bibinfo {author} {\bibfnamefont {A.}~\bibnamefont
  {Bartschi}}\ and\ \bibinfo {author} {\bibfnamefont {S.}~\bibnamefont
  {Eidenbenz}},\ }\bibfield  {title} {\bibinfo {title} {Short-depth circuits
  for dicke state preparation},\ }in\ \href
  {https://doi.org/10.1109/qce53715.2022.00027} {\emph {\bibinfo {booktitle}
  {2022 IEEE International Conference on Quantum Computing and Engineering
  (QCE)}}}\ (\bibinfo  {publisher} {IEEE},\ \bibinfo {year} {2022})\ p.\
  \bibinfo {pages} {87–96}\BibitemShut {NoStop}%
\bibitem [{\citenamefont {Liu}\ \emph {et~al.}(2025{\natexlab{d}})\citenamefont
  {Liu}, \citenamefont {Childs},\ and\ \citenamefont
  {Gottesman}}]{liu2025lowdepthquantumsymmetrization}%
  \BibitemOpen
  \bibfield  {author} {\bibinfo {author} {\bibfnamefont {Z.}~\bibnamefont
  {Liu}}, \bibinfo {author} {\bibfnamefont {A.~M.}\ \bibnamefont {Childs}},\
  and\ \bibinfo {author} {\bibfnamefont {D.}~\bibnamefont {Gottesman}},\ }\href
  {https://arxiv.org/abs/2411.04019} {\bibinfo {title} {Low-depth quantum
  symmetrization}} (\bibinfo {year} {2025}{\natexlab{d}}),\ \Eprint
  {https://arxiv.org/abs/2411.04019} {arXiv:2411.04019 [quant-ph]} \BibitemShut
  {NoStop}%
\bibitem [{\citenamefont {Yuan}\ and\ \citenamefont
  {Zhang}(2025)}]{yuan2025depthefficientquantumcircuitsynthesis}%
  \BibitemOpen
  \bibfield  {author} {\bibinfo {author} {\bibfnamefont {P.}~\bibnamefont
  {Yuan}}\ and\ \bibinfo {author} {\bibfnamefont {S.}~\bibnamefont {Zhang}},\
  }\href {https://arxiv.org/abs/2505.15413} {\bibinfo {title} {Depth-efficient
  quantum circuit synthesis for deterministic dicke state preparation}}
  (\bibinfo {year} {2025}),\ \Eprint {https://arxiv.org/abs/2505.15413}
  {arXiv:2505.15413 [quant-ph]} \BibitemShut {NoStop}%
\bibitem [{\citenamefont {Fannes}\ \emph {et~al.}(1992)\citenamefont {Fannes},
  \citenamefont {Nachtergaele},\ and\ \citenamefont
  {Werner}}]{fannes1992finitely}%
  \BibitemOpen
  \bibfield  {author} {\bibinfo {author} {\bibfnamefont {M.}~\bibnamefont
  {Fannes}}, \bibinfo {author} {\bibfnamefont {B.}~\bibnamefont
  {Nachtergaele}},\ and\ \bibinfo {author} {\bibfnamefont {R.~F.}\ \bibnamefont
  {Werner}},\ }\bibfield  {title} {\bibinfo {title} {Finitely correlated states
  on quantum spin chains},\ }\href@noop {} {\bibfield  {journal} {\bibinfo
  {journal} {Communications in mathematical physics}\ }\textbf {\bibinfo
  {volume} {144}},\ \bibinfo {pages} {443} (\bibinfo {year}
  {1992})}\BibitemShut {NoStop}%
\end{thebibliography}%
\end{document}

\title{Supplemental Material for ``Lovász meets Lieb-Schultz-Mattis: Complexity in Approximate Quantum Error Correction"}
	\newcommand{\PI}{Perimeter Institute for Theoretical Physics, Waterloo, Ontario, Canada N2L 2Y5}
	\newcommand{\UW}{Department of Physics and Astronomy, University of Waterloo, Waterloo, Ontario, Canada N2L 3G1}
	\newcommand{\Dalhousie}{Department of Mathematics and Statistics, Dalhousie University, Halifax, Nova Scotia, Canada B3H 4R2}
	\newcommand{\IBM}{IBM Quantum, IBM Research}
	
	\author{Jinmin Yi}
	\affiliation{\PI}
	\affiliation{\UW}
	
	\author{Ruizhi Liu}
	\affiliation{\Dalhousie}
	\affiliation{\PI}
	
	\author{Zhi Li}
	\affiliation{\IBM}

	\maketitle
	\tableofcontents\

	In this supplemental material, we provide more details related to the main text. 
	Specifically, in Sec.~\ref{sec:LLL} we review the Lovász local lemma (LLL) and provide a proof for a generalized version of the Lopsided Lovász local lemma. 
	In Sec.~\ref{sec:maintheorem_supp}, we provide the detailed proofs for the two main theorems in the main text and establish a more general theorem. 
	In Sec.~\ref{sec:clustering}, we discuss the scope of our results, showing that the conditions of the theorem(s) are satisfied by several classes of quantum states, including normal matrix product states.
	In Sec.~\ref{sec:covariantcode_supp}, we provide the detailed proofs for the code subspace complexity lower bound for covariant codes. 
	In Sec.~\ref{sec:Wstate_supp}, we prove the complexity lower bound for W states. 
	In Sec.~\ref{sec:LSM_supp}, we prove two Lieb-Schultz-Mattis type theorems under our framework.
	
	\section{Extensions and proof of the Lovász local lemma}\label{sec:LLL}
	
	In this section, we review the Lovász local lemma and present a generalization and its proof. 
	
	The Lovász local lemma states that if we have a collection of \emph{mostly independent} events and each event has a small enough individual probability, then there exists a positive lower bound on the probability that none of these events occur. 
	
	Let $\{A_i\}$, where $i\in [n]:=\{1,2,\cdots,n\}$ be a set of events.
	For each $i$, we assign a set of ``adjacent'' labels, denoted by $\Gamma(i)\subseteq [n]-{i}$.
	We first present the following symmetric version of the Lovász local lemma (with a concrete lower bound), which is used in the main text.
	We prove it later as a corollary of a more general theorem.
	\begin{theorem}\label{thm:LLL}
		Suppose for $\forall i\in [n]$ we have:
		\begin{itemize}
			\item each $A_i$ is independent from the collection $\{A_j|j\in [n]-\{i\}-\Gamma(i)\}$\footnote{Note that this is stronger than $A_i$ being independent with each $A_j$ where $j\in [n]-\{i\}-\Gamma(i)\}$.};
			\item $|\Gamma(i)|\leq d$ and $\mathbb{P}(A_1)\leq p\leq \frac{1}{e(d+1)}$, where $e=exp(1)$ is the Euler's number;
		\end{itemize}
		then   
		\begin{equation}
			\mathbb{P}\left(\overline{A_1} \wedge \cdots \wedge \overline{A_n}\right) >(1-ep)^n.
		\end{equation}
	\end{theorem}
	
	While the standard LLL assumes strict independence outside a dependency neighborhood, correlations in quantum many-body systems typically decay with distance rather than vanish. 
	This motivates the following generalization of LLL, where we relax the strict independence condition to be that events that are not neighbors of $A$ have limited influence on $\mathbb{P}(A)$, as described in \cref{eq-glll-condition1}.
	
	\begin{theorem}[Generalized Lopsided Lovász local lemma]\label{thm:GLLLL}
		Suppose there exists a constant $c\geq 1$ and an assignment of real numbers $x_i\in[0,1/c)$ to the events, such that for all $i\in[n]$, we have:
		\begin{equation}\label{eq-glll-condition1}
			\mathbb{P}(A_i|\wedge_{j\in S}\overline{A_j})\leq c\mathbb{P}(A_i),
			~~\forall \mathcal{S}\subseteq [n]-\Gamma(i)-\{i\}
		\end{equation}
		and
		\begin{equation}\label{eq-glll-condition2}
			\mathbb{P}(A_i)\leq x_i\prod_{j\in \Gamma(i)}(1-cx_j).
		\end{equation}
		Then
		\begin{equation}\label{eq-glll}
			\mathbb{P}\left(\overline{A_1} \wedge \cdots \wedge \overline{A_n}\right) \geq \prod_{i=1}^n\left(1-cx_i\right).
		\end{equation}
	\end{theorem}
	
	\begin{proof}
		We claim that:
		\begin{equation}\label{eq:ind}
			\mathbb{P}\left(A_i | \wedge_{j\in S}\overline{A_j}\right)\le c\,x_i \text{~whenever~} i\in[n], S\subseteq [n], i\notin S.
		\end{equation}
		Assuming \cref{eq:ind}, the conclusion \cref{eq-glll} follows by the chain rule:
		\begin{equation}
			\mathbb{P}\!\left(\wedge_{i=1}^n \overline{A_i}\right)
			=\prod_{i=1}^n \mathbb{P}\!\left(\overline{A_i} | \wedge_{j<i}\overline{A_j}\right)
			\ge \prod_{i=1}^n \bigl(1-c\,x_i\bigr),
		\end{equation}
		where we used \cref{eq:ind} with \(S=\{1,\dots,i-1\}\) to get
		\(\mathbb{P}(\overline{A_i} \mid \wedge_{j<i}\overline{A_j})\ge 1-c\,x_i\).
		
		We prove \cref{eq:ind} by induction on \(|S|\). The base case is \(|S|=0\). 
		By \cref{eq-glll-condition2},
		\begin{equation}
			\mathbb{P}(A_i)\le x_i\prod_{j\in \Gamma(i)}\bigl(1-c\,x_j\bigr)\le x_i\le c\,x_i\;,
		\end{equation}
		since $c\geq 1$ and each factor in the product does not exceed 1. 
		Thus \cref{eq:ind} holds if $|S|=0$.
		
		Now we proceed with induction. For a given non-empty set $S$, assume that \cref{eq:ind} holds for any set with size smaller than $|S|$.
		Split $S$ into $S_1:=S\cap \Gamma(i)$ and $S_2:=S\setminus S_1$. 
		If $S_1=\varnothing$, then $\mathcal{S}\subseteq [n]-\Gamma(i)-\{i\}$. Combining \cref{eq-glll-condition1} and \cref{eq-glll-condition2}, one obtains
		\begin{equation}\label{eq:S_1=0case}
			\mathbb{P}\left(A_i | \wedge_{j\in S}\overline{A_j}\right)\leq c\mathbb{P}(A_i)\leq cx_i\prod_{j\in \Gamma(i)}(1-cx_j)\leq cx_i\;,
		\end{equation}
		Hence \cref{eq:ind} holds.

		If $S_1\neq \varnothing$, then
		\begin{equation}\label{eq:ratio}
			\mathbb{P}\left(A_i | \wedge_{j\in S}\overline{A_j}\right)
			=\frac{\mathbb{P}\left(A_i \wedge (\wedge_{j\in S_1}\overline{A_j} )\middle| \wedge_{j\in S_2}\overline{A_j}\right)}
			{\mathbb{P}\left(\wedge_{j\in S_1}\overline{A_j} | \wedge_{j\in S_2}\overline{A_j}\right)}
			\leq
			\frac{\mathbb{P}\left(A_i | \wedge_{j\in S_2}\overline{A_j}\right)}
			{\mathbb{P}\left(\wedge_{j\in S_1}\overline{A_j} | \wedge_{j\in S_2}\overline{A_j}\right)} .
		\end{equation}
		The numerator can be bounded similarly to \cref{eq:S_1=0case}: since \(S_2 \subseteq [n]\setminus\bigl(\Gamma(i)\cup\{A_i\}\bigr)\), combining  \cref{eq-glll-condition1} and \cref{eq-glll-condition2}, one obtains
		\begin{equation}\label{eq:numerator}
			\mathbb{P}\left(A_i | \wedge_{j\in S_2}\overline{A_j}\right)\leq c\,\mathbb{P}(A_i)
			\leq c\,x_i\prod_{j\in \Gamma(i)}\!\!\bigl(1-c\,x_j\bigr).
		\end{equation}
		For the denominator, denoting $S_1=\{{j_1},\dots,{j_r}\} (r\geq 1)$, by the chain rule and the induction hypothesis (applied to each \(j_t\) with the conditioning set \(S_2\cup\{A_{j_1},\dots,A_{j_{t-1}}\}\), which has size \(<|S|\)),
		\begin{equation}\label{eq:denominator}
			\mathbb{P}\!\left(\wedge_{j\in S_1}\overline{A_j} | \wedge_{j\in S_2}\overline{A_j}\right)
			=\prod_{t=1}^r \mathbb{P}\!\left(\overline{A_{j_t}} |( \wedge_{\ell<t}\overline{A_{j_\ell}})\wedge (\wedge_{j\in S_2}\overline{A_j})\right)
			\geq \prod_{j\in S_1}\bigl(1-c\,x_j\bigr). 
		\end{equation}
		Combining \cref{eq:ratio} with \cref{eq:numerator} and \cref{eq:denominator} gives
		\[
		\mathbb{P}\!\left(A_i | \wedge_{j\in S}\overline{A_j}\right)
		\le c\,x_i\!\!\prod_{j\in \Gamma(i)\setminus S_1}\!\!\bigl(1-c\,x_j\bigr)
		\le c\,x_i,
		\]
		since each factor in the product does not exceed 1. 
		This completes the induction and the proof.
	\end{proof}
	
	There are two corollaries following from the above theorem.
	The first one is the usual ``asymmetric LLL", obtained by setting $c=1$, copied below for convenience.
	\begin{corollary}[Asymmetric LLL]
		Suppose there exists an assignment of real numbers $x_i\in[0,1)$ to the events, such that for all $i\in[n]$, we have:
		\begin{itemize}
			\item each $A_i$ is independent from the collection $\{A_j|j\in [n]-\{i\}-\Gamma(i)\}$;
			\item $\mathbb{P}(A_i)\leq x_i\prod_{j\in \Gamma(i)}(1-x_j)$;
		\end{itemize}
		then
		\begin{equation}
			\mathbb{P}\left(\overline{A_1} \wedge \cdots \wedge \overline{A_n}\right) \geq \prod_{i=1}^n\left(1-x_i\right).
		\end{equation}
	\end{corollary} 
	
	The second corollary is a generalized version of the symmetric LLL.
	In particular, taking $c=1$, we recover \cref{thm:LLL}.
	\begin{corollary}\label{thm:GSLLL}
		Assuming there exists a constant $c\geq 1$, and that for $\forall i\in [n]$ we have:
		\begin{itemize}
			\item \cref{eq-glll-condition1} holds;
			\item $|\Gamma(i)|\leq d$ and $\mathbb{P}(A_i)\leq p\leq \frac{1}{ce(d+1)}$, where $e=exp(1)$;
		\end{itemize}
		then   
		\begin{equation}
			\mathbb{P}\left(\overline{A_1} \wedge \cdots \wedge \overline{A_n}\right) >(1-cep)^n.
		\end{equation}
	\end{corollary}
	\begin{proof}
		Consider the function $f(x)=x(1-cx)^d$. It is increasing for $x\in[0,\frac{1}{c(d+1)}]$ and decreasing for $x\in[\frac{1}{c(d+1)},1/c]$. 
		Since
		\begin{equation}
			f(\frac{1}{c(d+1)})=\frac{1}{c(d+1)}(1-\frac{1}{d+1})^d > \frac{1}{ce(d+1)} \geq p,
		\end{equation}
		there must exist an $x_0\in (0,\frac{1}{c(d+1)})$ such that $f(x_0)=p$.
		Moreover, since
		\begin{equation}
			f(x_0)=x_0(1-cx_0)^d>x_0(1-\frac{1}{d+1})^d>\frac{x_0}{e},
		\end{equation}
		we have $x_0<ep$.

		Let us apply \cref{thm:GLLLL}.
		We have assumed that \cref{eq-glll-condition1} holds.
		Now set $x_i=x_0$, we get
		\begin{equation}
			x_i\prod_{j\in \Gamma(i)}(1-cx_j)\geq f(x_0)=p \geq \prob{A_i},
		\end{equation}
		so the condition \cref{eq-glll-condition2} is also satisfied.
		Therefore, \cref{thm:GLLLL} implies that 
		\begin{equation}
			\mathbb{P}\left(\overline{A_1} \wedge \cdots \wedge \overline{A_n}\right) \geq 
			\prod_{i=1}^n\left(1-cx_0\right)^n > (1-cep)^n.
		\end{equation}
	\end{proof}

	\section{Extensions and detailed proof of the main theorems}\label{sec:maintheorem_supp}
	
	In this section, we present a general statement and the complete proof for the distinguishability theorem for short-range correlated states, as shown in \cref{thm:SRC_Distinguishability}. 
	As a special case, it leads to the distinguishability \cref{thm_supp:local-distinguishability} for short-range entangled states as in the main text. 
	We then prove the incompatibility between the AQEC capability and short-range entanglement, as shown in \cref{Thm_supp:SV_LowerBound}.
	
	\subsection{General result}
	
	\begin{theorem}\label{thm:SRC_Distinguishability}
		Let $\mathcal{G}=\prod_{i=1}^n (1-P_i)$ be the projector to the ground state subspace of a commuting projector Hamiltonian $H=\sum_{i=1}^n P_i$. Suppose for a (possibly mixed) quantum state $\rho$, there exists a constant $c\geq 1$, and a family of regions $\{R_i\}_{i\in[n]}$, such that:
		\begin{itemize}
			\item for any positive semidefinite operator $Q$, whenever $\mathrm{supp}(Q)\cap R_i=\varnothing$, we have
			\begin{equation}\label{eq:clustercondition}
				\langle P_i Q\rangle_{\rho} 
				\leq c\braket{P_i}_\rho\braket{Q}_\rho;
			\end{equation}
			\item for all $i$, $\mathrm{supp}(P_j)\cap R_i\neq \varnothing$ for at most $K$ numbers of $j\neq i$;
			\item $ce(K+1)p\leq 1$, where $p=\max_{i}\braket{P_i}_\rho$ and $e=\exp(1)$.
		\end{itemize}
		Then $\tr(\mathcal{G}\rho)> (1-cep)^n$.
	\end{theorem}
	\begin{proof}
		Consider the probability distribution defined by simultaneously measuring $P_i$ on $\rho$.
		Let $A_i$ represent the event in which measuring $P_i$ yields a result of 1. By definition, $\prob{A_i}=\braket{P_i}_\rho$. For each $i$, define the dependency neighborhood
		\begin{equation}
			\Gamma_i\coloneqq \{j\neq i|\mathrm{supp}(P_j)\cap R_i\neq \varnothing\}.
		\end{equation}
		Then \cref{eq:clustercondition} implies that:
		\begin{equation}
			\mathbb{P}(A_i|\land_{j\in S}\overline{A}_j)\leq c\mathbb{P}(A_i),~~\forall S\subseteq [n]\backslash(\Gamma_i\cup\{i\}).
		\end{equation}
		To see this, let $Q=\prod_{j\in S} (1-P_j)$. Since $\mathrm{supp}(Q)\subseteq \cup_{j\in S} \mathrm{supp}(P_j)$ and $S\cap(\Gamma_i\cup\{i\})=\varnothing$, we have $\mathrm{supp}(Q)\cap R_i=\varnothing$, satisfying the condition for \cref{eq:clustercondition}.
		
		Now we can \cref{thm:GSLLL}, which implies
		\begin{equation}
			\mathbb{P}\left(\wedge_{i=1}^n\overline{A_i}\right) > (1-cep)^n\;.
		\end{equation}
		Since $\tr(\mathcal{G}\rho) = \mathbb{P}\left(\wedge_{i=1}^n\overline{A_i}\right)$, this completes the proof.
	\end{proof}

	\subsection{Corollaries: theorems in the main text}\label{sec:2B}
	We will discuss condition~\cref{eq:clustercondition} in more detail in \cref{sec:MPSclustering_supp}. For now, let us note the following. 
	Suppose a quantum state $\ket{\psi}$ is prepared by a depth-$t$ circuit $U$, i.e., $\ket{\psi} = U\ket{0^n}$. Consider two operators $P$ and $Q$.
	If $\supp(U^\dagger PU)\cap \supp(U^\dagger QU)=\varnothing$, then
	\begin{equation}
		\bra{\psi}PQ\ket{\psi}=\bra{0^n}U^\dagger PU U^\dagger QU\ket{0^n} = \bra{0^n}U^\dagger PU \ket{0^n}\bra{0^n} U^\dagger QU\ket{0^n} = \bra{\psi}P\ket{\psi} \bra{\psi}Q\ket{\psi}.
	\end{equation}
	Namely, \cref{eq:clustercondition} holds with equality and $c=1$.
	Moreover, a sufficient condition for $\supp(U^\dagger PU)\cap \supp(U^\dagger QU)=\varnothing$ is that $\supp(Q)$ does not intersect the depth-$2t$ light cone of $\supp(P)$\footnote{Here the light cone is defined by stacking circuit structure of $U^\dagger$ and $U$ (we only use the circuit structure, so we cannot use $U^\dagger U=\mathbbm{1}$).
		This light cone may depend on the circuit (especially in the all-to-all case), but this is not a problem.}.
	
	With this in mind, now we show that \cref{thm:SRC_Distinguishability} implies the distinguishability theorem for short-range entangled (SRE) states in the main text. 
	
	\begin{theorem}[Distinguishability of SRE states]
		\label{thm_supp:local-distinguishability}
		Let $|\psi_{1,2}\rangle$ be two $n$-qubit states that are almost orthogonal, \ie $|\langle\psi_1|\psi_2\rangle|\leq\delta$. Suppose $\cc(|\psi_{1,2}\rangle)\leq t$, then there exists a distinguishing operator $\mO$, with $\opnorm{\mO}= 1$ and $|\mathrm{supp}(\mO)| \leq f(t)$ (here $f(t)$ is the light cone function), such that
		\begin{equation}
			|\langle\psi_1|\mO|\psi_1\rangle-\langle\psi_2|\mO|\psi_2\rangle|>\frac{2}{e}\min\{1-\delta^{\frac{2}{n}},\frac{1}{f(4t)}\}.
			\label{eq:distinguishability-bound}
		\end{equation}
	\end{theorem}
	
	\begin{proof}
		Since $\cc(\ket\psi_{1,2})\leq t$, they can be written as
		\begin{align}
			|\psi_{1,2}\rangle&=U_{1,2}|0^n\rangle,
		\end{align}
		with $\mathrm{depth}(U_{1,2})\leq t$. 
		Thus $|\psi_{1}\rangle$ is the unique ground state of a commuting projector Hamiltonian $H^{(1)}=\sum_{i=1}^n P_i^{(1)}$, with $P^{(1)}_i=(1-U_{1}ZU_{1}^\dagger)/2$.
		% So does $|\psi_{2}\rangle$ with $H^{(2)}=\sum_{i=1}^n P_i^{(2)}=\sum_{i=1}^n (1-U_{2}ZU_{2}^\dagger)/2$.
		We denote $\mathcal{G}=\prod_j(1-P_j)=|\psi_1\rangle\langle\psi_1|$. 
		
		We apply \cref{thm:SRC_Distinguishability} to the Hamiltonian $H^{(1)}=\sum_{i=1}^n P_i^{(1)}$ and the state $\ket{\psi_2}$.
		We take $R_i$ to be the depth-$2t$ light cone of $\supp(P_i^{(1)})$.
		As discussed above, $\braket{P_i^{(1)}Q}_{\psi_2}=\braket{P_i^{(1)}}_{\psi_2}\braket{Q}_{\psi_2}$ whenever $\supp(Q)\cap R_i=\varnothing$.
		Moreover, since each $\supp(P_i^{(1)})$ itself is contained in the depth-$t$ light cone of site $i$,
		we have that, for all $i$, $\mathrm{supp}(P_j^{(1)})\cap R_i\neq \varnothing$ for at most $f(4t)-1$ values of $j\neq i$.

		Denote $p=\max_i\langle\psi_2|P_i^{(1)}|\psi_2\rangle$. 
		By \cref{thm:SRC_Distinguishability}, we either have
		\begin{equation}
			p>\frac{1}{ef(4t)},
		\end{equation}
		or
		\begin{equation}
			\tr(\mathcal{G}|\psi_2\rangle\langle\psi_2|)>(1-ep)^n.
		\end{equation}
		Note that $\tr(\mathcal{G}|\psi_2\rangle\langle\psi_2|)=|\langle\psi_1|\psi_2\rangle|^2\leq\delta^2$, we thus have
		\begin{equation}
			p>\min\{\frac{1-\delta^{\frac{2}{n}}}{e},\frac{1}{ef(4t)}\}\;.
		\end{equation}
		Pick $P_*^{(1)}$ to be the projector where we get the maximum in $\max_i\langle\psi_2|P_i^{(1)}|\psi_2\rangle$, 
		and define $\mO=2P_*^{(1)}-1$, we have $\opnorm{\mO}=1$, $|\supp(\mO)|\leq f(t)$, and
		\begin{equation}
			|\langle\psi_1|\mO|\psi_1\rangle-\langle\psi_2|\mO|\psi_2\rangle|
			=2|\langle\psi_1|P_*^{(1)}|\psi_1\rangle-\langle\psi_2|P_*^{(1)}|\psi_2\rangle|
			=2\langle\psi_2|P_*^{(1)}|\psi_2\rangle
			>\frac{2}{e}\min\{1-\delta^{\frac{2}{n}},\frac{1}{f(4t)}\}.
		\end{equation}
	\end{proof}
	
	\begin{theorem}
		\label{Thm_supp:SV_LowerBound}
		Given an $(\!(n,k)\!)$ quantum code with two orthogonal code states $|\psi_1\rangle$ and $|\psi_2\rangle$, if $\cc^\delta(|\psi_{1(2)}\rangle)\leq t$ with
		\begin{equation}
			\delta\leq\left(1-\frac{1}{f(4t)}\right)^{\frac{n}{2}},
		\end{equation}
		then
		\begin{equation}
			\varepsilon(f(t))>\frac{1}{ef(4t)}-\delta.
		\end{equation}
		Here $f(t)$ denotes the lightcone function for a given connectivity constraint.
	\end{theorem}
	\begin{proof}
		Since there are two code states $|\psi_{1(2)}\rangle$ with approximate complexity upper bound $\cc^\delta(|\psi_{1(2)}\rangle)\leq t$, there exist two states two states $|\psi'_1\rangle$ and $|\psi'_2\rangle$, in the $\delta$-vicinity of $|\psi_1\rangle$ and $|\psi_2\rangle$ respectively, that can be written as
		\begin{align}
			|\psi'_1\rangle&=U_1|0^n\rangle\;;\\
			|\psi'_2\rangle&=U_2|0^n\rangle\;,
		\end{align}
		with $\cc(U_{1(2)})\leq t$. 
		Consider the Fubini–Study angle $d_{FS}(\psi,\phi)\coloneqq \arccos |\braket{\psi|\phi}|$, note that
		\begin{equation}
			\onenorm{|\psi\rangle\langle\psi|-|\phi\rangle\langle\phi|}=2\sqrt{1-|\braket{\psi|\phi}|^2}=2\sin d_{FS}(\psi,\phi)\;.
		\end{equation}
		
		Using the triangle inequality, we have
		\begin{equation}
			\begin{split}
				|\braket{\psi_1'|\psi_2'}|&=\cos(d_{FS}(\psi'_1,\psi_2'))\leq \cos(\pi/2-d_{FS}(\psi'_1,\psi_1)-d_{FS}(\psi'_2,\psi_2))\\
				&\leq \sin(2\arcsin\frac{\delta}{2})\leq \delta\;.
			\end{split}
		\end{equation}
		
		Thus, by~\cref{thm_supp:local-distinguishability}, if 
		\begin{equation}
			\delta\leq\left(1-\frac{1}{f(4t)}\right)^{\frac{n}{2}}\;,
		\end{equation}
		then there exists some operator $\mO$ such that $\supp(\mO)\leq f(t)$ and
		\begin{equation}
			|\langle\psi'_1|\mO|\psi'_1\rangle-\langle\psi'_2|\mO|\psi'_2\rangle|>\frac{2}{ef(4t)}.
		\end{equation}
		
		For simplicity of notation, we denote $R=\supp(\mO)$. 
		Consider the difference of reduced density matrix on the region $R$, \ie $\delta\rho'_R=\tr_{\overline{R}}(|\psi'_1\rangle\langle\psi'_1|-|\psi'_2\rangle\langle\psi'_2|)$.
		Using H\"older’s inequality, we deduce
		\begin{equation}
			|\langle\psi'_1|\mO|\psi'_1\rangle-\langle\psi'_2|\mO|\psi'_2\rangle|=\tr\left[\delta\rho'_RO_k\right]\leq \onenorm{\delta\rho'_RO_k}\leq\onenorm{\delta\rho'_R}\opnorm{O_k}=\onenorm{\delta\rho'_R}.
		\end{equation}
		Using the triangle inequality, we obtain
		\begin{equation}
			\begin{split}
				\onenorm{\delta\rho'_R}&=\onenorm{(\psi'_{1R}-\Gamma_R)-(\psi'_{2R}-\Gamma_R)}\\
				&\leq \onenorm{\psi_{1R}-\Gamma_R}+\onenorm{\psi_{2R}-\Gamma_R}+\onenorm{\psi_{1R}-\psi'_{1R}}+\onenorm{\psi_{2R}-\psi'_{2R}}\leq2\varepsilon(f(t))+2\delta,
			\end{split}
		\end{equation}
		where we used $\psi_R$ to denote the reduced density matrix of $|\psi\rangle$ on region $R$.
		
		Thus, we have proved
		\begin{equation}
			\varepsilon(f(t))>\frac{1}{ef(4t)}-\delta.
		\end{equation}
	\end{proof}

	\section{Scope of the theorem}\label{sec:clustering}
	
	In this section, we discuss examples where \cref{eq:clustercondition} holds. 
	\Cref{eq:clustercondition} can be viewed as a clustering condition, stating that distinct regions do not exhibit strong correlations.
	We will show that the following classes of states satisfy this condition:
	\begin{enumerate}
		\item pure states prepared by a finite-depth circuit from product states;
		\item unique ground states of commuting projector Hamiltonians;
		\item mixed states prepared by a finite-depth quantum channel from product states;
		\item logical states (pure or mixed) of commuting projector LDPC codes such that each $\supp(P_i)$ is correctable.
		\item normal matrix product states (MPS).
	\end{enumerate}
	For cases 1-4, $c=1$. For case 5, $c>1$ may happen.

	\subsection{States with zero correlation length}
	
	\textbf{Case 1.}
	We have already proved it at the beginning of \cref{sec:2B}.
	Alternatively, note that any pure state prepared by a finite-depth circuit $|\psi\rangle=U|0^n\rangle$ is also a unique ground state of the commuting projector Hamiltonian $H=\sum (1- UZ_iU^\dagger)/2$, 
	so case 1 is also a special instance of case 2.
	
	\textbf{Case 2.}
	Consider the unique ground state $|\psi\rangle$ of a commuting projector Hamiltonian $H=\sum P'_i$. 
	For any two observables $P$ and $Q$ that are distant enough such that no projector $P'_j$ can intersect their support simultaneously, we can define 
	\begin{equation}
		\Pi_{\check{P}}=\prod_{\mathrm{supp}(P'_j)\cap\mathrm{supp}(P)=\varnothing}(1-P'_j)\;,
	\end{equation}
	and $\Pi_{\check{Q}}$ similarly.
	We have $[P,\Pi_{\check P}]=[Q,\Pi_{\check Q}]=0$, $\Pi_{\check P}\Pi=\Pi_{\check Q}\Pi=\Pi_{\check P}\Pi_{\check Q}=\Pi$,  hence
	\begin{equation}
		\langle\psi|PQ|\psi\rangle=\langle\psi|\Pi P\Pi_{\check P}\Pi_{\check Q}Q\Pi|\psi\rangle
		=\langle\psi|\Pi P\Pi Q\Pi|\psi\rangle=\langle\psi|P|\psi\rangle\langle\psi|Q|\psi\rangle\;.
	\end{equation}
	Thus \cref{eq:clustercondition} is automatically satisfied with $c=1$ if we take $R_i$ to be the union of $\supp(P'_j)$ for all $P'_j$ whose support intersects with $\mathrm{supp}(P_i)$.
	
	\textbf{Case 3.}
	We can always regard a finite-depth quantum channel as a finite-depth unitary circuit followed by tracing out the ancilla.
	Thus, case 3 reduces to case 1, and 
	we only need to take $R_i$ to cover the depth-$2t$ light cone of $\mathrm{supp}(P_i)$.
	
	\textbf{Case 4.}
	The proof is similar to that of case 2.
	Consider a quantum code defined by commuting projectors $P_i'$.
	The code space projector is $\Pi=\prod (1-P_{i}')$.
	Any logical state $\rho$, pure or mixed, satisfies $\rho \Pi=\Pi\rho=\rho$.
	For any two observable $P$ and $Q$ such that no $P_i'$ intersects with both of their supports, we define $\Pi_{\check P}$ and $\Pi_{\check Q}$ similarly.
	Then 
	\begin{equation}
		\tr(\rho PQ)=\tr(\rho\Pi P\Pi_{\check P}\Pi_{\check Q}Q\Pi)
		=\tr(\rho\Pi P\Pi Q\Pi)
		=\tr(\rho P)\tr(\rho Q).
	\end{equation}
	Here in the last equality we used the Knill-Laflamme condition with the assumption that $\mathrm{supp}(P)$ is correctable, and thus $\Pi P\Pi=\tr(\rho P)\Pi$. 
	A sufficient but not necessary condition is $|\supp(P)|<d$, where $d$ is the code distance.
	
	Thus \cref{eq:clustercondition} is automatically satisfied with $c=1$ if we take $R_i$ to contain the support of every $P_j'$ whose support intersects with $\mathrm{supp}(P_i)$.
	Note that, while this conclusion is general, it is particularly useful for an LDPC code, in which case $\supp(R_i)$ will be bounded by $\supp(P_i)$ and a locality parameter.

	\subsection{Clustering property for normal matrix product state}\label{sec:MPSclustering_supp}
	
	In this subsection, we provide a short review of matrix product states and then prove our clustering property \cref{eq:clustercondition} for normal MPS. 
	
	%\begin{align}
	%\ket{\psi[A]}&=
	% \begin{array}{c}
		% \begin{tikzpicture}[a/.style={fill=blue,fill opacity = .2},scale=.5,every node/.style={sloped,allow upside down},baseline={([yshift=-.8ex]current bounding box.center)}]
			% \def\a{1}
			% \def\dx{1}
			%     \foreach \x in {0,1,...,8}{
				%     \filldraw[ten,shift={(\x*\a+\x*\dx,0)}] (-\a/2,-\a/2) -- (-\a/2,\a/2) -- (\a/2,\a/2) -- (\a/2,-\a/2) -- (-\a/2,-\a/2);
				% \draw[shift={(\x*\a+\x*\dx,0)}] (-\dx/2-\a/2,0) -- (-\a/2,0);
				% \draw[shift={(\x*\a+\x*\dx,0)}] (\a/2,0) -- (\a/2+\dx/2,0);
				% \draw[shift={(\x*\a+\x*\dx,0)}] (0,-\a/2-\dx/2) -- (0,-\a/2);
				% }
			% \draw (-\dx/2-\a/2,0)--(-\dx/2-\a/2,\a)--(8.5\a+8.5\dx,\a)--(8.5\a+8.5\dx,0);
			%   \end{tikzpicture}.
		%   \end{array}
	% \end{align}
We focus on translationally invariant MPS. 
With periodic boundary condition, 
\begin{equation}\label{eq_supp:MPSdef}
	|\psi[A]\rangle=\sum_{i_1i_2...i_N}\tr[A_{i_1}A_{i_2}\cdots A_{i_N}]|i_1i_2...i_N\rangle\;.
\end{equation}
For any operator $\hat{O}$ with $|\supp(\mO)|=d$, we can define a $\hat{O}$-transfer matrix based on its matrix element $O_{\mathbf{ij}}=\langle\mathbf{i}|\mO|\mathbf{j}\rangle$ where $\mathbf{i}=(i_1,i_2,..., i_d)$ and $\mathbf{j}=(j_i,j_2,..., j_d)$ denote a basis for $d$-qubit quantum states. 
The $\hat{O}$-transfer matrix is then defined as
\begin{equation}
	\EE_\mO=\sum_{\mathbf{i},\mathbf{j}}O_{\mathbf{ij}}A_\mathbf{i}\otimes\bar{A}_\mathbf{j},
\end{equation}
where $A_\mathbf{i}=A_{i_1}A_{i_2} \cdots A_{i_d}$. 
If $\hat{O}$ is positive semi-definite, then $\EE_\mO$ is completely positive when viewed as a map (superoperator) $\text{Mat}(D)\to \text{Mat}(D)$, where $D$ is the bond dimension of the MPS.
In particular, if $\mO=\mathbbm{1}$, the 1-qubit identity operator, we just refer to $\EE_\mathbbm{1}=\sum_{i,j}A_i\otimes\bar{A}_j$ as the transfer matrix and denote it as $\EE$. 

An MPS is normal if $\EE$ has a unique eigenvalue of magnitude (and value) equal to 1, and the corresponding left and right eigenvectors are positive definite matrices \cite{cirac2021matrix}.
We denote a second largest (in magnitude) eigenvalue as $\lambda_2$.
Furthermore, we can perform a gauge transformation so that the right eigenvector is the identity matrix $\mathbbm{1}$.
We denote the left eigenvector as $\rho$:
\begin{equation}
	\EE(\mathbbm{1})=\sum_i A_iA_i^\dagger=\mathbbm{1},~~
	\EE^\dagger(\rho)=\sum_i A_i^\dagger \rho A_i=\rho,~~
	\tr(\rho)=1.
\end{equation}

Now we prove the clustering property \cref{eq:clustercondition} for normal MPS on the infinite chain. 
The cases of open boundary conditions and periodic boundary conditions can be proved similarly.

The correlation function between two spatially local positive semi-definite operators $\hat{P}$ and $\hat{Q}$ in the thermodynamic limit can be expressed in terms of the transfer matrix as:
\begin{align}
	\langle\psi[A]|\hat{P}|\psi[A]\rangle 
	&= \tr(\rho\EE_{\hat{P}}\mathbbm{1}),\\
	\langle\psi[A]|\hat{Q}|\psi[A]\rangle 
	&= \tr(\rho\EE_{\hat{Q}}\mathbbm{1}), \\
	\langle\psi[A]|\hat{P}\hat{Q}|\psi[A]\rangle
	&= \tr(\rho\EE_{\hat{P}}\EE^{\ell}\EE_{\hat{Q}}\mathbbm{1}) \notag
	=\tr(\rho\EE_{\hat{P}}\EE^{\infty}\EE_{\hat{Q}}\mathbbm{1})
	+ \tr(\rho\EE_{\hat{P}}(\EE^{\ell}-\EE^\infty)\EE_{\hat{Q}}\mathbbm{1}) \notag\\
	&= \tr(\rho\EE_{\hat{P}}\mathbbm{1})\,
	\tr(\rho\EE_{\hat{Q}}\mathbbm{1})
	+ \tr(\rho\EE_{\hat{P}}(\EE^{\ell}-\EE^\infty)\EE_{\hat{Q}}\mathbbm{1}).
\end{align}
Here $\ell$ denotes the distances between $\supp(\hat{P})$ and $\supp(\hat{Q})$.
Note that for any $\lambda$ such that $|\lambda_2| < \lambda <1$,  we have
\begin{equation}\label{eq:specapprox}
	|\tr(\rho\EE_{\hat{P}}(\EE^{\ell}-\EE^\infty)\EE_{\hat{Q}}\mathbbm{1})|
	\leq \lambda^\ell \Fnorm{\EE_{\hat{Q}}\mathbbm{1}} \Fnorm{\EE_{\hat{P}}^\dagger\rho}
\end{equation}
for $\ell$ large enough (depending on $\lambda$ and $D$).
Here $\Fnorm{A}=\sqrt{\tr(A^\dagger A)}$ is the Frobenius norm, the norm induced by the natural inner product in $\text{Mat}(D)$.

Note that for a positive semi-definite matrix $A$ and a positive definite matrix $B$,
\begin{equation}
	\Fnorm{A}\leq \tr(AB)\sigma_{\max}(B^{-1}),
\end{equation}
where $\sigma_{\max}(B^{-1})$ denotes the maximal eigenvalue of $B^{-1}$. This can be seen by diagonalizing $A=a_j|j\rangle\langle j|$ and then
\begin{equation}
	\Fnorm{A}\leq \sum_j a_j\leq\sum_j a_j\langle j|B|j\rangle\sigma_{\max}(B^{-1})=\tr(AB)\sigma_{\max}(B^{-1}),
\end{equation}
where in the second inequality we use $\langle j|B|j\rangle\sigma_{\max}(B^{-1})\geq 1$.
In particular, since $\EE_{\hat{Q}}\mathbbm{1}$ and  $\EE_{\hat{P}}^\dagger\rho$ are positive definite, we have
\begin{equation}
	\begin{aligned}
		\Fnorm{\EE_{\hat{Q}}\mathbbm{1}}
		&\leq \tr(\rho\EE_{\hat{Q}}\mathbbm{1})\sigma_{\max}(\rho^{-1}),\\
		\Fnorm{\EE_{\hat{Q}}^\dagger\rho}
		&\leq \tr(\rho\EE_{\hat{P}}\mathbbm{1})\sigma_{\max}(\mathbbm{1})=\tr(\rho\EE_{\hat{P}}\mathbbm{1}).
	\end{aligned}
\end{equation}

We thus have
\begin{equation}
	\left|\tr(\rho\EE_{\hat{P}}(\EE^{\ell}-\EE^\infty)\EE_{\hat{Q}}\mathbbm{1})\right|
	\leq 
	\lambda^\ell \sigma_{\max}(\rho^{-1}) \tr(\rho\EE_{\hat{Q}}\mathbbm{1}) \tr(\rho\EE_{\hat{P}}\mathbbm{1}),
\end{equation}
and hence
\begin{equation}
	\langle\psi[A]|\hat{P}\hat{Q}|\psi[A]\rangle\leq\langle\psi[A]|\hat{P}|\psi[A]\rangle\langle\psi[A]|\hat{Q}|\psi[A]\rangle (1+\lambda^\ell\sigma_{\max}(\rho^{-1}))\;.
\end{equation}

To conclude, we can pick $\lambda \in (|\lambda_2|,1)$ and set $c = 1 + \lambda^\ell \sigma_{\max}(\rho^{-1})$, where $\ell$ is the smallest integer for which \cref{eq:specapprox} holds. Then \cref{eq:clustercondition} is satisfied for any positive semi-definite $\hat{P}$ and $\hat{Q}$ whose supports are separated by at least $\ell$.

\section{Complexity of covariant codes}\label{sec:covariantcode_supp}
In this section, we will review covariant codes, \ie error-correcting codes with transversal logical gates, and show that our results provide a complexity lower bound for their code states. 

We denote the encoding isometry as $V_{L\rightarrow A}$, which maps from an abstract logical space $L$ to a subspace of a physical space $A=A_1\otimes A_2\otimes \cdots \otimes A_n$. 
Let $G$ be a group that acts unitarily on the logical and physical systems as $U_L(g)$ and $U_A(g)$ respectively.
We require $U_A(g)$ to be transversal, \ie $U_A(g)=\bigotimes_iU_A^i(g)$. 
A code is covariant if
\begin{equation}
	V_{L\rightarrow A}U_L(g)=\left(\bigotimes_iU_A^i(g)\right)V_{L\rightarrow A}.
\end{equation}

\subsection{Universal gate sets}

If a code admits a universal set of transversal logical gates, then any logical gate can, by definition, be approximated to arbitrary accuracy by compositions of those gates, which remain transversal.
We then have the following corollary:
\begin{corollary}\label{thm:covariant_code_1}
	Given an $(\!(n,k)\!)$ covariant code with a universal transversal gate set, if 
	\begin{equation}\label{eq:covariant_code_condition}
		\varepsilon(f(t))\leq\frac{1}{ef(4t)}\;,
	\end{equation}
	then any code state will have a circuit complexity lower bound $\cc(\ket\psi)>t$.
\end{corollary}
Note that codes in this corollary are necessarily AQECs, due to the Eastin-Knill theorem.
\begin{proof}
	For any code state $\ket{\psi}$ and any $\delta>0$, there exists a transversal logical gate that takes $\ket{\psi}$ to an almost orthogonal state $\ket{\psi'}$ such that $|\braket{\psi|\psi'}|\leq \delta$.
	Since this logical gate is transversal, we have $\cc(|\psi\rangle)=\cc(|\psi'\rangle)$. 
	Now, assume by contradiction that $\cc(|\psi\rangle)\leq t$.
	Then by \cref{thm_supp:local-distinguishability}, we have 
	\begin{equation}
		\varepsilon(f(t)) > \frac{1}{e}\min\{1-\delta^{\frac{2}{n}},\frac{1}{f(4t)}\}.
	\end{equation}
	We can take $\delta$ to be arbitrarily small, hence $\varepsilon(f(t))>\frac{1}{ef(4t)}$, which contradicts \cref{eq:covariant_code_condition}. 
	Thus for any state $\ket{\psi}$, $\cc(\ket\psi)>t$.
\end{proof}

\Cref{thm:covariant_code_1} gives a complexity lower bound for the whole code subspace while being much weaker than the condition in~\cite{yi_complexity_2024}.

In the all-to-all case, $f(t)=2^{t}$, hence $f(4t)=f(t)^4$. Therefore, if there exists an $x$ such that $\varepsilon(x) \ll \frac{1}{x^4}$, we can then choose $t=f^{-1}(x)$ and get a complexity lower bound $\cc(\ket\psi)>t$.
If the largest $x$ satisfying the inequality is diverging with respect to $n$, the corresponding lower bound will be a superconstant.
For example, if $\varepsilon(x)\sim\frac{x}{n}$, we can pick $x=O(n^{\frac{1}{5}})$ with a small enough proportional constant. The corresponding lower bound will be $\cc(\ket\psi)=\Omega(\log n)$.

In the geometrically local case, $f(t)=(2t+1)^D$, hence $f(4t)\sim 4^D f(t)$.
Therefore, if there exists an $x$, preferably diverging with $n$,  such that $\varepsilon(x) \ll \frac{1}{x}$, we can still choose $t=f^{-1}(x)$ and get a complexity lower bound $\cc(\ket\psi)>t$. 
In the example of $\varepsilon(x)\sim\frac{x}{n}$, we may pick $x=O(n^{\frac{1}{2}})$ with a small enough proportional constant.
The corresponding lower bound will be $\cc(\ket\psi)=\Omega(n^\frac{1}{2D})$.

\subsection{Clifford gates}
Our results also enable us to obtain a code subspace complexity lower bound for any codes with transversal Clifford gates. 
\begin{corollary}
	Given an $(\!(n,k)\!)$ code with transversal Clifford logical gates, if 
	\begin{equation}
		\varepsilon(f(t))\leq \frac{1}{e}\min\left\{1-2^{-\frac{k}{n}},\frac{1}{f(4t)}\right\}\;,
	\end{equation}
	then any code state will have a circuit complexity lower bound $\cc(|\psi\rangle)>t$.
\end{corollary}

\begin{proof}
	For any state $\ket{\psi}$, if we randomly pick a Clifford logical gate $U\in\mathcal{C}_k$, since $\mathcal{C}_k$ forms a 2-design~\cite{Dankert2009}, we have
	\begin{equation}
		\frac{1}{|\mathcal{C}_k|}\sum_{U\in\mathcal{C}_k}|\langle\psi|U|\psi\rangle|^2=\int_{U(2^k)} dU |\langle\psi|U|\psi\rangle|^2=\frac{1}{2^k}\;.
	\end{equation}
	Therefore, it is evident that there must exist a Clifford logical gate $U$ such that $|\langle\psi|U|\psi\rangle|^2\leq 1/2^k$. Consequently, there exists a state $|\psi'\rangle$ in the code subspace such that $|\langle\psi|\psi'\rangle|^2\leq 1/2^k$.
	
	By \cref{thm_supp:local-distinguishability}, there exists a distinguishing operator $\mO$ with $\opnorm{\mO}=1$ and $|\supp(\mO)|\leq f(t)$ such that
	\begin{equation}
		|\langle\psi_1|\mO|\psi_1\rangle-\langle\psi_2|\mO|\psi_2\rangle|>\frac{2}{e}\min\{1-2^{-\frac{k}{n}},\frac{1}{f(4t)}\}.
	\end{equation}
	Thus, by H\"older's inequality and the triangular inequality, we obtain
	\begin{equation}
		\varepsilon(f(t))\geq \frac{1}{2}|\langle\psi_1|\mO|\psi_1\rangle-\langle\psi_2|\mO|\psi_2\rangle|>\frac{1}{e}\min\{1-2^{-\frac{k}{n}},\frac{1}{f(4t)}\}.
	\end{equation}
	Since we assumed the opposite, we have proved that there cannot be a code state $|\psi\rangle$ with $\cc(|\psi\rangle)\leq t$. 
\end{proof}

Note that the second part of the constraint is identical to the case of codes with a universal gate set. For the first part, if we are interested in a $t$ such that $\varepsilon(f(t)) \leq k/(2en)$, the condition is automatically satisfied.
This result is stronger than those previously known~\cite{yi_complexity_2024}, where the constraint on the subsystem variance included an additional logarithmic factor.

\subsection{Remark}
For completeness, let us remark that in the case where $G$ is a Lie group with a $U(1)$ subgroup that acts nontrivially, there exists a lower bound for the subsystem variance $\varepsilon=\Omega(1/n)$, making previous techniques in~\cite{yi_complexity_2024} inapplicable for lower bounding the complexity. 

To see this, let us expand the $U(1)$ action on both the logical and physical systems: 
\begin{equation}
	U_L(\theta)=e^{-i\theta T_L}\;,\quad U_A(\theta)=e^{-i\theta T_A}=\otimes_{i=1}^n e^{-i\theta T_i}\;,
\end{equation}
where $\theta\in [0,2\pi)$.
To study $\varepsilon(1)$, the subsystem variance for one qubit, note that 
\begin{equation}\label{eq_supp:DeltaTL}
	\Delta T_L=\max_{|\psi_{1,2}\rangle}\left(\langle\psi_1|T_A|\psi_1\rangle-\langle\psi_2|T_A|\psi_2\rangle\right)=\max_{|\psi_{1,2}\rangle}\sum_{i=1}^{n}\left(\langle\psi_1|T_i|\psi_1\rangle-\langle\psi_2|T_i|\psi_2\rangle\right)\leq 
	n\varepsilon(1)\max_i\Delta T_i,
\end{equation}
where $\Delta T_{i}$ denotes the range of the eigenvalues of $T_{i}$, and $\Delta T_{L}$ denotes the range of $\braket{T_L}$ within the logical space, which we have assumed to be nontrivial.
Therefore,
\begin{equation}
	\varepsilon(1)\geq\frac{\Delta T_L}{n\max_i\Delta T_i}.
\end{equation}

\section{W state preparation}\label{sec:Wstate_supp}

In this section, we review the complexity bounds for the $n$-qubit $W$ state, both in the geometrically local and all-to-all cases. 
Recall that the $W$ state is defined as follows:
\begin{equation}\label{eq:Wstate}
	|W_n\rangle \;=\; \frac{1}{\sqrt{n}}\sum_{i=1}^n |0\cdots 01_i0\cdots 0\rangle\;.
\end{equation}

\subsection{Proof via local indistinguishability}
\begin{corollary}\label{thm:CofW}
	For $\delta< 1/10$, the geometrically local circuit complexity of $|W_n\rangle$ on a 1D chain is 
	\begin{equation}
		\cc^\delta(|W_n\rangle)=\Omega(n).
	\end{equation}
	For $\delta<1/n^\alpha$ with $\alpha>1/2$, the all-to-all circuit complexity of $|W_n\rangle$ is 
	\begin{equation}
		\cc^\delta(|W_n\rangle)=\Omega(\log n).
	\end{equation}
\end{corollary}
\begin{proof}
	Note that $|W_n\rangle$ is approximately locally indistinguishable from the product state $|0^n\rangle$, thus if $|W_n\rangle$ is short-range entangled, \cref{thm_supp:local-distinguishability} will be violated. 
	
	Concretely, for the geometric complexity, we divide the $n$ qubits into patches of size $m$. 
	The value $m$ is to be chosen later.
	Suppose $\cc^\delta(|W_n\rangle)\leq t$, then there exists some state $|\psi\rangle=U|0^n\rangle$ with $\mathrm{depth}(U)\leq t$, such that $\onenorm{|\psi\rangle\langle\psi|-|W_n\rangle\langle W_n|}\leq \delta$. 
	We now apply \cref{thm:SRC_Distinguishability} to the state $|\psi\rangle$, by taking $P_i=1-|0^m\rangle\langle0^m|$ where $i$ is the index for different patches. We thus have
	\begin{equation}
		\max_i\left|\langle\psi|P_i|\psi\rangle\right|\leq \frac{\delta}{2}+\max_i\left|\langle W_n|P_i|W_n\rangle\right|=\frac{\delta}{2}+\frac{m}{n}.
	\end{equation}
	To satisfy \cref{eq:clustercondition}, we again take $R_i$ to be the lightcone of $\supp(P_i)$ under a depth-$2t$ circuit, \ie $|R_i|=m+4t$, in which case  \cref{eq:clustercondition} will be saturated with $c=1$. And for all $i$, $\mathrm{supp}(P_j)\cap R_i\neq \varnothing$ for at most $|R_i|/m+1$ values of $j\neq i$. By \cref{thm:SRC_Distinguishability}, if
	\begin{equation}\label{eq_supp:Wcondition}
		(\frac{m+4t}{m}+2)(\frac{\delta}{2}+\frac{m}{n})\leq \frac{1}{e}\;,
	\end{equation}
	then we would have
	\begin{equation}
		\tr (\mathcal{G}|\psi\rangle\langle\psi|)>(1-e(\frac{\delta}{2}+\frac{m}{n}))^{\frac{n}{m}}.
	\end{equation}
	On the other hand, 
	\begin{equation}
		\tr (\mathcal{G}|\psi\rangle\langle\psi|)=|\langle\psi|0^n\rangle|^2
		\leq \sin^2(\theta(\psi,W_n))
		=\frac{\delta^2}{4}.   
	\end{equation}
	Now we take $m=\delta n$.
	Therefore, if
	\begin{equation}
		(1-\frac{3e}{2}\delta)^{1/\delta}\geq \frac{\delta^2}{4}\;,
	\end{equation}
	which can be numerically solved to be satisfied when $\delta\leq 0.1207$, \cref{eq_supp:Wcondition} should be reversed. Picking $\delta=1/10$, we get $t>n/3$ and thus $\cc^\delta(|W_n\rangle)=\Omega(n)$.
	
	For the all-to-all case, we proceed similarly. However, in this case, $|R_i|=2^{2t} m$; for each $i$, $\mathrm{supp}(P_j)\cap R_i\neq \varnothing$ holds for at most $(2^{2t} m-1)$ values of $j\neq i$. 
	By \cref{thm:SRC_Distinguishability}, if
	\begin{equation}\label{eq_supp:Wcondition_all2all}
		(m\cdot2^{2t})(\frac{\delta}{2}+\frac{m}{n})\leq \frac{1}{e}\;,
	\end{equation}
	then $\tr (\mathcal{G}|\psi\rangle\langle\psi|)>(1-e(\frac{\delta}{2}+\frac{m}{n}))^{n/m}$. 
	
	For simplicity, we again take $m/n=\delta$.
	Therefore, if $(1-\frac{3e}{2}\delta)^{1/\delta}\geq \frac{\delta^2}{4}$, which can be satisfied if we take $\delta=n^{-\alpha}$ with $\alpha>1/2$, then \cref{eq_supp:Wcondition_all2all} must be violated, \ie
	\begin{equation}
		2^{2t-1}n^{1-2\alpha}>\frac{1}{3e}\;,
	\end{equation}
	thus $t=\Omega(\log n)$.
\end{proof}

\subsection{Proof via long-range correlation}
In this subsection, we prove the circuit complexity lower bound for the W state using a different method.
The key observation is that W states have long-range correlations.
\begin{corollary_restate}{thm:CofW}{$'$}
	For $\delta< 1/3$, the geometrically local circuit complexity of $|W_n\rangle$ on a 1D chain is 
	\begin{equation}
		\cc^\delta(|W_n\rangle)=\Omega(n).
	\end{equation}
	For $\delta<1/n^\alpha$ with $\alpha>0$, the all-to-all circuit complexity of $|W_n\rangle$ is 
	\begin{equation}
		\cc^\delta(|W_n\rangle)=\Omega(\log n).
	\end{equation}  
\end{corollary_restate}
\begin{proof}
	For any two subregions $A$ and $B$ such that $|A|=|B|=k\leq \frac{n}{2}$, we claim that:
	\begin{equation}
		\opnorm{W_{AB}-W_A\otimes W_B}_1 > \frac{2k}{n},
	\end{equation}
	where $W_{A}$ is the reduced density matrix of the W state on region $A$ (similarly for $B$ and $AB$).
	Indeed, notice that $W_{A}\otimes W_{B}$ has spectra $\{(\frac{k}{n})^2,\frac{k}{n}(1-\frac{k}{n}),\frac{k}{n}(1-\frac{k}{n}),(1-\frac{k}{n})^2\}$ and $W_{AB}$ has the spectra $\{\frac{2k}{n},1-\frac{2k}{n}\}$.
	Therefore, due to the Hoffman-Wielandt inequality, we have\footnote{Explicit calculation shows that the equality holds.}:
	\begin{equation}
		\opnorm{W_{AB}-W_A\otimes W_B}_1 
		\geq |(1-\frac{k}{n})^2-(1-\frac{2k}{n})|+|\frac{k}{n}(1-\frac{k}{n})-\frac{2k}{n}|+\frac{k}{n}(1-\frac{k}{n})+(\frac{k}{n})^2
		=\frac{2k}{n}+\frac{2k^2}{n^2}.
	\end{equation}
	As a result, for any state $\psi$ such that $\opnorm{\psi-W}_1\leq \epsilon \leq \frac{2k}{3n}$, triangle inequality then implies that
	\begin{equation}
		\psi_{AB}\neq \psi_A\otimes\psi_B.
	\end{equation}
	
	Now consider the geometric local case.
	Given the target error $\delta$, we pick $k=\frac{3\delta}{2}n$.
	Suppose $\psi$ is a pure state such that $\opnorm{\psi-W}_1<\delta$, then $t=\cc(\ket\psi)$ must satisfies
	\begin{equation}
		k+2t> \frac{n}{2}.
	\end{equation}
	Otherwise, we can always pick $A$ and $B$ such that they are $2t$ away from each other, and the standard light cone argument implies that $\psi_{AB}=\psi_A\otimes\psi_B$.
	Therefore, if $\delta<\frac{1}{3}$, we have $t=\Omega(n)$.

	For the all-to-all case, the arguments are similar.
	If $\psi$ is prepared by a depth-$t$ circuit and $(2^{2t}+1)k\leq n$, a light cone argument shows that there exists $A$ and $B$ such that $|A|=|B|=k$ and $\psi_{AB}=\psi_A\otimes\psi_B$, a contradiction.
	Therefore, for $\delta=n^{-\alpha}~(\alpha>0)$, then we can pick $k=\Theta(n^{1-\alpha})$ and deduce that $t=\Omega(\log(n^\alpha))=\Omega(\log(n))$.
\end{proof}

\section{Lieb-Schultz-Mattis type theorems}\label{sec:LSM_supp}
This section reviews the Lieb–Schultz–Mattis (LSM) constraint and presents a detailed proof, along with a proof of its momentum descendant.

\subsection{$U(1)\times T$ LSM}
Consider a 1D system of size $L$, with the lattice translation operator $\hat{T}$ and an on-site $U(1)$-symmetry
\begin{equation}
	U_\theta=e^{i\theta\hat{Q}}=\bigotimes_{x=0}^{L-1} e^{i\theta\hat{q}_x}\;,
\end{equation}
where $\theta\in [0,2\pi)$ is the $U(1)$-angle, $\hat{q}_x$ is the local charge operator on site $x$ with only integer eigenvalues and $\hat T\hat{q}_x \hat{T}^\dagger=\hat{q}_{x+1}$ and $\hat{Q}=\sum_{x=0}^{L-1}\hat{q}_x$ is the total charge. 
Note that $||\hat{q}||=||\hat{q}_x||$ is independent of $x$, since $||\hat{q}_{x+1}||=||\hat{T}\hat{q}_{x}\hat{T}^{\dagger}||$.

\begin{theorem}[$U(1)\times T$ LSM]\label{thm_supp:LSM}
	Suppose $\ket{\psi}$ is both translationally invariant
	\begin{align}
		\hat{T}\ket{\psi}\propto \ket{\psi},
	\end{align}
	and $U(1)$ symmetric with non-commensurate charge filling
	\begin{equation}\label{eq:non-commensurate}
		\exp(\frac{2\pi i\hat{Q}}{L})\ket{\psi} = e^{i\alpha} \ket{\psi},~~e^{i\alpha}\neq 1.
	\end{equation}
	Then if $\cc^\delta(|\psi\rangle)=t$, we have either
	\begin{equation}\label{eq_supp:lsm_cond1}
		\delta> \left(1-e\delta-e(\frac{9\pi t^2\opnorm{\hat{q}}}{2L})^2\right)^{L/t}\;,
	\end{equation}
	or
	\begin{equation}\label{eq_supp:lsm_cond2}
		t\geq L^{1/2}\sqrt{\frac{2}{9\pi\opnorm{\hat{q}}}(\frac{1}{7e}-\delta)}\;.
	\end{equation}
\end{theorem}

\begin{proof}
	Assume that the geometric local complexity $\cc^\delta(|\psi\rangle)=t$, \ie there exists a state $|\phi\rangle=V|0^L\rangle$ that is $\delta$-close to $|\psi\rangle$ with $\mathrm{depth}(V)=t$ ($\ket{\phi}$ is not necessarily translationally invariant). 
	Now we can consider the large gauge transformation
	\begin{equation}
		U:=\exp(\frac{2\pi i}{L}\sum_{x=0}^{L-1}x\,\hat{q}_{x})\;,
	\end{equation}
	and we consider the state $|\tilde{\psi}\rangle=U|\psi\rangle$. If $|\psi\rangle$ has momentum $p$, \ie
	\begin{equation}
		\hat{T}|\psi\rangle=e^{ip}|\psi\rangle\;,
	\end{equation}
	then $|\tilde{\psi}\rangle$ has a different momentum $p-\alpha\not=p\,(\text{mod}\,2\pi)$, since:
	\begin{equation}
		\begin{split}
			\hat{T}|\tilde{\psi}\rangle&=\hat{T}\exp(\frac{2\pi}{L}\sum_{x=0}^{L-1}x\,\hat{q}_{x})|\psi\rangle=\exp(\frac{2\pi}{L}\sum_{x=0}^{L-1}x\,\hat{q}_{x+1})\hat{T}|\psi\rangle\\
			&=U\exp(-\frac{2\pi i\hat{Q}}{L})e^{ip}\ket{\psi}=e^{i(p-\alpha)}|\tilde{\psi}\rangle\;.
		\end{split}
	\end{equation}
	Thus we have $\langle\psi|\tilde{\psi}\rangle=0$. 
	
	Now consider $|\phi\rangle=V|0^L\rangle$ and $|\tilde{\phi}\rangle=UV|0^L\rangle$, which are $\delta$-close to $|\psi\rangle$ and $|\tilde{\psi}\rangle$, respectively. By the triangular inequality of the Fubini-Study metric, we have
	\begin{equation}
		|\braket{\phi|\tilde{\phi}}|\leq \sin (2\arcsin\frac{\delta}{2})\leq \delta\;.
	\end{equation}
	Since $|\phi\rangle$ and $|\tilde{\phi}\rangle$ are unique ground states of commuting projector models, we can now apply \cref{thm:SRC_Distinguishability} to the state $|\tilde{\phi}\rangle$, by taking $P_i=V(1-|0^m\rangle\langle0^m|)V^\dagger$ where $i$ is the index for different patches, with the dependency number $K=\frac{4t}{m}+2$. Taking $p=\max_i \left|\langle\tilde{\phi}|P_i|\tilde{\phi}\rangle\right|$, by \cref{thm:SRC_Distinguishability}, we either have
	\begin{equation}\label{eq:lsm_proof_cond1}
		(\frac{4t}{m}+3)p>\frac{1}{e},
	\end{equation}
	or
	\begin{equation}\label{eq:lsm_proof_cond2}
		|\braket{\phi|\tilde{\phi}}|>(1-ep)^{L/m}.
	\end{equation}

	We now estimate $p$. Consider $P_i$ for an arbitrary $i$.
	Denote $s=|\supp(P_i)|\leq m+2t$.
	There exists an $\ell\in\mathbb{Z}$, such that $\supp(P_i)\subseteq \ell+S$, where $S=[-\frac{s-1}{2},\frac{s-1}{2}]$ (if $m$ odd) or $S=[-\frac{s}{2}+1,\frac{s}{2}]$ (if $m$ even). 
	
	Since $\ket{\psi}$ is invariant under $\exp(\frac{2\pi i}{L}\hat{Q})$, $\ket{\phi}$ is approximately invariant: for $\forall\ell\in\mathbb{Z}$,
	\begin{align}
		\opnorm{ \exp(\frac{-2\pi i\ell}{L}\hat{Q})\ket{\phi}\bra{\phi}\exp(\frac{2\pi i\ell}{L}\hat{Q}) - \ket{\phi}\bra{\phi} }_1 \leq 2\delta.
	\end{align}
	Applying $U$, we get:
	\begin{equation}\label{eq:LSM1}
		\opnorm{ U_\ell\ket{\phi}\bra{\phi}U_\ell^\dagger - \ket{\tilde\phi}\bra{\tilde\phi} }_1 \leq 2\delta,
	\end{equation}
	where
	\begin{equation}
		U_\ell=U\exp(\frac{-2\pi i\ell}{L}\hat{Q}) =\exp(\frac{2\pi i}{L}\sum_{x=0}^{L-1}x\hat{q}_{x+\ell}).
	\end{equation}
	Due to the tensor product structure of $U_\ell$, we have
	\begin{equation}\label{eq:LSM2}
		\bra{\phi}U_\ell^\dagger P_i U_\ell\ket{\phi} = \bra{\phi}U_S^\dagger P_i U_S\ket{\phi},
	\end{equation}
	where $U_S=\exp(\frac{2\pi i}{L}\sum_{x\in S}x \hat{q}_{x+\ell})$.
	Note that for any Hermitian operator $A$, we have
	\begin{equation}
		\opnorm{e^{iA}-1} = \max_{a\in \text{spec}(A)}|e^{i a}-1| 
		\leq \opnorm{A}.
	\end{equation} 
	Therefore,
	\begin{equation}
		\opnorm{U_S-1} \leq \sum_{x\in S} \opnorm{\exp(\frac{2\pi i}{L} x \hat{q}_{x+\ell})-1}
		\leq \sum_{x\in S} \opnorm{\frac{2\pi i}{L} x \hat{q}_{x+\ell}}
		\leq \frac{\pi s^2}{2L}\opnorm{\hat{q}}\;.
	\end{equation}
	Using $P_i\ket{\phi}=0$, we have
	\begin{equation}\label{eq:LSM3}
		\bra{\phi}U_S^\dagger P_i U_S\ket{\phi}     
		=\opnorm{P_iU_S\ket{\phi}}^2
		=\opnorm{P_i(U_S-1)\ket{\phi}}^2
		\leq\opnorm{U_S-1}^2
		\leq \left(\frac{\pi s^2}{2L}\opnorm{\hat{q}}\right)^2.
	\end{equation}
	Consequently, combining \cref{eq:LSM1,eq:LSM2,eq:LSM3}, we get:
	\begin{equation}
		\bra{\tilde\phi} P_i \ket{\tilde\phi}
		\leq    \delta+ \bra{\phi}U_\ell^\dagger P_i U_\ell\ket{\phi} 
		\leq \delta+ \left(\frac{\pi (m+2t)^2}{2L}\opnorm{\hat{q}}\right)^2,
	\end{equation}
	which is an upper bound for $p$.

	Combining \cref{eq:lsm_proof_cond1,eq:lsm_proof_cond2} and taking $m=t$, we get that either
	\begin{equation}
		\delta> \left(1-e\delta-e(\frac{9\pi t^2\opnorm{\hat{q}}}{2L})^2\right)^{L/t}\;,
	\end{equation}
	or
	\begin{equation}
		t\geq L^{1/2}\sqrt{\frac{2}{9\pi\opnorm{\hat{q}}}(\frac{1}{7e}-\delta)}.
	\end{equation}
\end{proof}

As corollaries, we derive results on the exact complexity of states under LSM constraints and on the impossibility of preparing these states via finite-time Hamiltonian evolution.
\begin{corollary}\label{corr_supp:LSM}
	Suppose $\ket{\psi}$ is both translationally invariant
	\begin{align}
		\hat{T}\ket{\psi}\propto \ket{\psi},
	\end{align}
	and $U(1)$ symmetric with non-commensurate charge filling
	\begin{equation}\label{eq:non-commensurate2}
		\exp(\frac{2\pi i\hat{Q}}{L})\ket{\psi} = e^{i\alpha} \ket{\psi},~~e^{i\alpha}\neq 1.
	\end{equation}
	Then 
	\begin{enumerate}[label=(\arabic*)]
		\item $\cc(|\psi\rangle)=\Omega(L^{1/2})$.
		\item $|\psi\rangle$ cannot be prepared from a product state by finite time Hamiltonian evolution.
	\end{enumerate}
\end{corollary}
\begin{proof}
	(1) Take $\delta$=0.
	
	(2) The proof is based on a circuit simulation of finite-time evolution. For any finite-time evolution $U_T^H$ on a 1D lattice of length $L$, it is known~\cite{HaahEvolutionCircuit} that there exists a quantum circuit $U_C$ with depth $O(\mathrm{polylog}(L/\delta))$, such that $\opnorm{U_C-U_T^H}\leq \delta$. 
	
	Assuming the contrary, that $|\psi\rangle$ can be prepared from a product state by finite time Hamiltonian evolution, \ie $|\psi\rangle=U_T^H|0^L\rangle$, consider the state $|\psi'\rangle=U_C|0^L\rangle$, we have
	\begin{equation}
		\onenorm{|\psi'\rangle\langle\psi'|-|\psi\rangle\langle\psi|} \leq 2\opnorm{\ket{\psi}-\ket{\psi}'}
		\leq 2\opnorm{U_C-U_T^H}\leq 2\delta.
	\end{equation}
	Therefore $\cc^{2\delta}(|\psi\rangle)=O(\mathrm{polylog}(L/\delta))$. 
	Taking $\delta=1/L$, then $t=\cc^{2\delta}(|\psi\rangle)=O(\mathrm{polylog} L)$.
	By \cref{thm_supp:LSM}, we have either \cref{eq_supp:lsm_cond1}, which requires that
	\begin{equation}
		\frac{1}{L}>(1-O(\frac{1}{L})-O(\frac{t^4}{L^2}))^{L/t}=O(e^{-1/t}),
	\end{equation}
	or
	\cref{eq_supp:lsm_cond2}, which requires that
	\begin{equation}
		t=\Omega(L^{1/2}).
	\end{equation}
	Neither relation can be satisfied.
	Thus $|\psi\rangle$ cannot be prepared from a product state by finite time Hamiltonian evolution.
\end{proof}

\subsection{Nonzero momentum implies LRE}

\begin{corollary}
	Let $|\psi\rangle$ be a state on a 1D system of size L. Suppose it is translationally invariant with a non-zero momentum:
	\begin{equation}
		\hat{T}|\psi\rangle=e^{ip}|\psi\rangle,\quad e^{ip}\neq 1,
	\end{equation}
	then the geometric local complexity $\cc(|\psi\rangle)=\Omega(L)$.
\end{corollary}
\begin{proof}

	\begin{figure}
		\centering
		\includegraphics[width=0.9\linewidth]{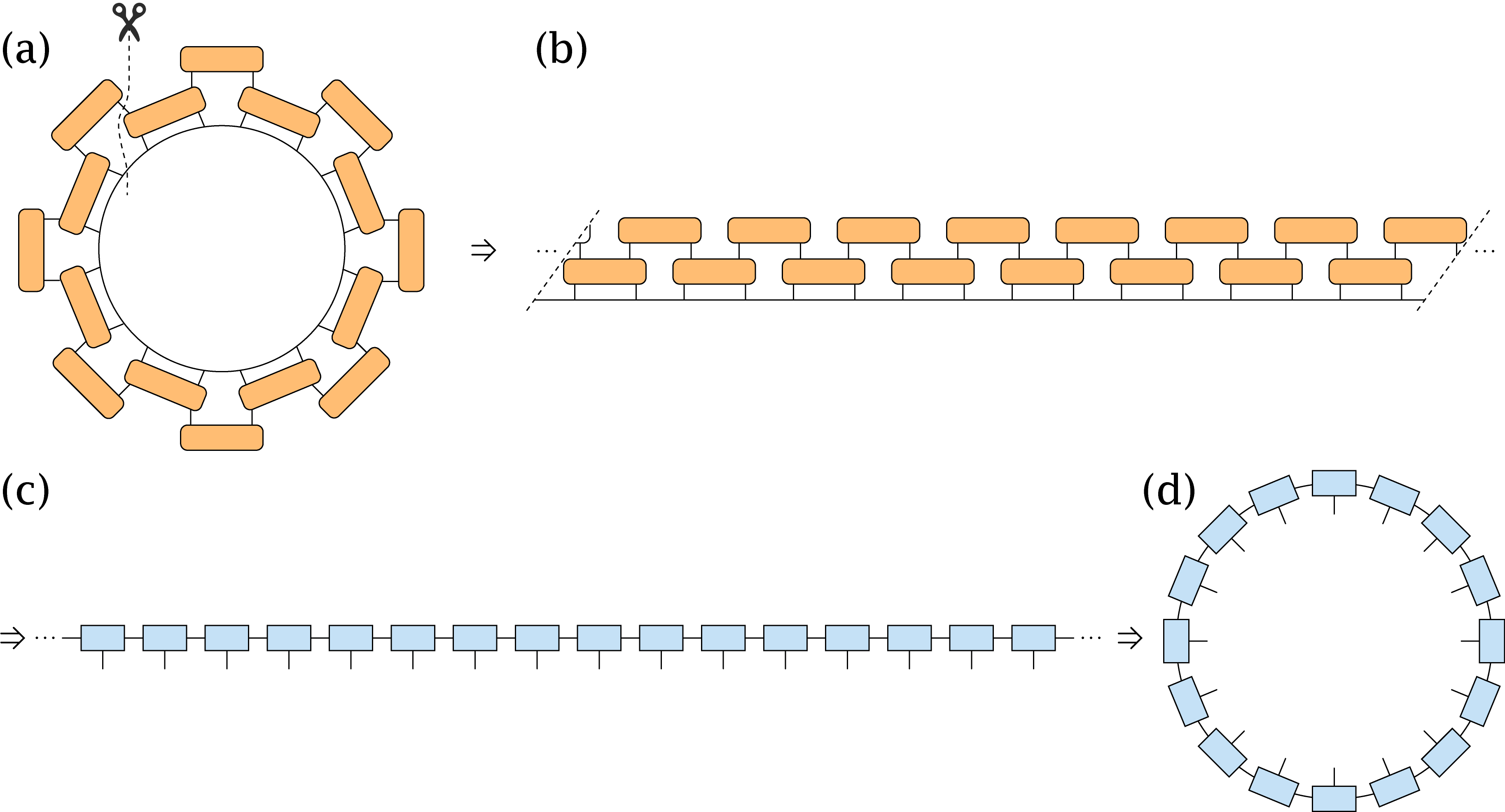}
		\caption{The construction of a partner state $|\psi'\rangle$ from the original state $|\psi\rangle =U|0^n\rangle$. (a) Split the circuit $U$ along an arbitrary vertical cut. (b) Replicate and tile the cut piece and obtain an infinite-lattice circuit $\tilde{U}$. (c) The resulting translational symmetric states generated by $\tilde{U}$ can be represented by an iMPS. (d) Truncating this iMPS yields $|\psi'\rangle$.}
		\label{fig:momentum}
	\end{figure}
	Suppose $\cc(|\psi\rangle)=t$, \ie $|\psi\rangle=U |0^n\rangle$ with $\mathrm{depth}(U)=t<n/4$. 
	While $U$ is defined on a period chain, we can extend it to a circuit $\tilde{U}$ on an infinite chain by cutting $U$ along an arbitrary vertical cut, making infinite copies, and pasting them as in Fig.~\ref{fig:momentum} (a), (b). 
	Formally, take $U=\prod_t U_t$ where $U_t$ is a layer of unitaries on a periodic lattice, we define $\tilde{U}=\prod_t \tilde{U}_t$ where each $\tilde{U}_t=U_t\otimes U_t\otimes U_t\otimes\cdots$ is the infinite extension of  $U_t$. 
	Denote $\omega_0$ to be the products of $|0\rangle\langle0|$ on the infinite chain, we consider the state 
	\begin{equation}
		\omega=\tilde{U}\omega_0\tilde{U}^\dagger.
	\end{equation}
	
	We claim that $\omega$ is translationally invariant. To see this, denote $\omega'=\hat T \omega \hat T^\dagger$ where $\hat T$ is the lattice translation. 
	Due to the construction of $\tilde{U}$ and the translational invariance of $|\psi\rangle$, we know that
	\begin{equation}
		\omega(A)=\bra{\psi}A\ket{\psi}=\bra{\hat{T}\psi}A\ket{\hat{T}\psi}=\omega'(A)
	\end{equation}
	for any operator $A$ such that $|\supp(A)|< n-2t$ (hence $\tilde{U}^\dagger A\tilde{U}=U^\dagger AU$).
	On the other hand, since $\omega_0$ is the unique ground state of the Hamiltonian $H=\sum_i T^i P T^{\dagger i}$ where $P=1-\ket{0}\bra{0}$ is a projector at site 0,
	$\omega$ is therefore the unique ground state of $\tilde{H}=\tilde{U}H\tilde{U}^\dagger$.
	Note that $\tilde{H}$ is $2t$-local, thus for each term in $\tilde{H}$, the expectation value should be the same for $\omega$ and $\omega'$, thus $\omega'$ is also a ground state of $\tilde{H}$. By uniqueness of the ground state, we thus have $\omega'=\omega$.
	
	Moreover, since $\omega$ can be generated by a product state from a finite depth circuit, it is finitely correlated in the sense of Ref.~\cite{fannes1992finitely}.
	It is also a pure state $C^*$-algebraic sense.
	Ref.~\cite{fannes1992finitely} (sec.~2-4) shows that such a state can always be purely generated with trivial peripheral spectra.
	In other words, $\omega$ can be represented as a translationally invariant infinite matrix product state (iMPS) as in Fig.~\ref{fig:momentum} (c), such that
	\begin{itemize}[itemsep=0pt, topsep=0pt]
		\item 1 is the only (and non-degenerated) eigenvalue of modulus 1 of the transfer matrix $\EE$;
		\item the right eigenvector of $\EE$ is $\mathbbm{1}$, where $\mathbbm{1}$ is the identity matrix;
		\item the left eigenvector of $\EE$ is $\rho$, where $\rho$ is positive definite.
	\end{itemize}
	Crucially, no coarse-graining is needed at this step.
	The above conditions also imply that the iMPS is injective after enough coarse-graining (see sec.~5 in Ref.~\cite{fannes1992finitely}).
	
	Due to the construction of $\omega$, for any two operators $\hat{P}$ and $\hat{Q}$ with $\mathrm{dist}(\supp(\hat{P}),\supp(\hat{Q}))=\ell>2t$,
	we have: 
	\begin{equation}
		\langle\hat{P}\hat{Q}\rangle_\omega
		=\langle\hat{P}\rangle_\omega\langle\hat{Q}\rangle_\omega.
	\end{equation}
	On the other hand, using the MPS structure, we have
	\begin{equation}
		\langle\hat{P}\hat{Q}\rangle_\omega - \langle\hat{P}\rangle_\omega\langle\hat{Q}\rangle_\omega
		=\tr(\rho\EE_{\hat{P}}(\EE^{\ell}-\EE^{\infty})\EE_{\hat{Q}}\mathbbm{1}).
	\end{equation}
	Here, the support of $\hat P$ and $\hat Q$ can be arbitrarily large, and they are not necessarily semi-definite.
	Now we fix the size of $\hat P$ and $\hat Q$ to be a large enough number $s$ such that the MPS is injective after coarse-graining $s$ physical sites into one.
	It follows that $\EE_{\hat Q}\mathbbm{1}$ and $\EE_{\hat Q}^\dagger\rho$ can be arbitrary matrices.
	Therefore, it follows that
	\begin{equation}\label{eq:Fcluster}
		\EE^\ell=\ket{\mathbbm{1}}\bra{\rho},~~\forall \ell>2t.
	\end{equation}
	Here $\ket{\mathbbm{1}}\bra{\rho}$ is the superoperator $X\mapsto \tr(\rho X)\mathbbm{1}$.
	
	Now we construct a state $\ket{\psi'}$ on $n$ qudits by truncating the iMPS and imposing the periodic boundary condition as in Fig.~\ref{fig:momentum} (d).
	Namely, 
	\begin{equation}    
		\ket{\psi'}=|\psi[A]\rangle=\sum_{i_1i_2...i_N}\tr[A_{i_1}A_{i_2}\cdots A_{i_N}]|i_1i_2...i_N\rangle\;.
	\end{equation}
	It is automatically translationally invariant with zero momentum: $\hat T\ket{\psi'}=\ket{\psi'}$.
	By \cref{eq:Fcluster}, for any operator $\hat{O}$ such that $|\supp(
	\hat O)|\leq n-2t$, we have
	\begin{equation}
		\bra{\psi'}\hat O\ket{\psi'} 
		= \text{Tr}(\EE^{L-|\supp(\hat O)|}\EE_{\hat O})
		=\tr(\rho\EE_{\hat O}\mathbbm{1})=        
		\omega(A) = \bra{\psi}\hat O\ket{\psi}.
	\end{equation}
	Here, Tr denotes the trace for superoperators.
	
	If $n-2t\geq 2t$, we can choose $\hat O$ to be $U(T^iPT^{-i})U^\dagger$, which collectively fixes $\ket{\psi}$ as the unique ground state.
	Therefore, $\ket{\psi'}$ and $\ket{\psi}$ are in fact the same state (up to a phase)\footnote{Alternatively, one may apply \cref{thm:SRC_Distinguishability} to deduce $|\langle\psi'|\psi\rangle|>0$.}.
	However, since $|\psi\rangle$ has non-zero momentum and $|\psi'\rangle$ has zero momentum, they must be orthogonal, a contradiction.
	We thus proved that
	\begin{equation}
		\cc(|\psi\rangle)> \frac{n}{4}.
	\end{equation}
\end{proof}

\bibliography{ref.bib}